\documentclass{aastex631}

\shorttitle{Simulated galaxy observations from NIHAO}
\shortauthors{Faucher et al.}

\graphicspath{{./}{figures/}}
\newcommand\HII{$\textrm{H}\scriptstyle\mathrm{II}$}
\begin{document}

\title{Panchromatic simulated galaxy observations from the NIHAO project}

\author{Nicholas Faucher}
\affiliation{Center for Cosmology and Particle Physics, Department of Physics, New York University,
726 Broadway, New York, New York 10003, USA}

\author{Michael R. Blanton}
\affiliation{Center for Cosmology and Particle Physics, Department of Physics, New York University,
726 Broadway, New York, New York 10003, USA}

\author{Andrea V. Macci\`{o}}
\affiliation{New York University Abu Dhabi, PO Box 129188, Abu Dhabi, United Arab Emirates}
\affiliation{Center for Astro, Particle and Planetary Physics (CAP$^{3}$), New York University Abu Dhabi, Abu Dhabi, United Arab Emirates}
\affiliation{Max-Planck-Institut f\"{u}r Astronomie, K\"{o}nigstuhl 17, D-69117 Heidelberg, Germany}

\begin{abstract}
\noindent 
We present simulated galaxy spectral energy distributions (SEDs) from the far ultraviolet through the far infrared, created using hydrodynamic simulations and radiative transfer calculations, suitable for the validation of SED modeling techniques. SED modeling is an essential tool for inferring star formation histories from nearby galaxy observations, but is fraught with difficulty due to our incomplete understanding of stellar populations, chemical enrichment processes, and the non-linear, geometry dependent effects of dust on our observations. Our simulated SEDs will allow us to assess the accuracy of these inferences against galaxies with known ground truth. To create the SEDs, we use simulated galaxies from the NIHAO suite and the radiative transfer code SKIRT. We explore different sub-grid post-processing recipes, using color distributions and their dependence on axis ratio of galaxies in the nearby universe to tune and validate them. We find that sub-grid post-processing recipes that mitigate limitations in the temporal and spatial resolution of the simulations are required for producing FUV to FIR photometry that statistically reproduce the colors of galaxies in the nearby universe. With this paper we release resolved photometry and spatially integrated spectra for our sample galaxies, each from a range of different viewing angles. Our simulations predict that there is a large variation in attenuation laws among galaxies, and that from any particular viewing angle that energy balance between dust attenuation and reemission can be violated by up to a factor of 3. These features are likely to affect SED modeling accuracy.
\end{abstract}

\section{Introduction} \label{Introduction}

\noindent Understanding our place in the universe requires
untangling the history of galaxies over cosmic time.
This paper presents simulated observations of galaxies 
accounting for dust absorption, scattering, and emission,
to test the spectral energy distribution (SED) modeling
techniques that many investigators use to infer the growth and evolution of galaxies from observations. 

Galaxy evolution research proceeds on two major, complementary 
avenues---the archaeological record of individual galaxies at 
redshift zero and look-back studies of how the population of 
galaxies changes over time. This paper concentrates on 
understanding the archaeological record. 

Unraveling galaxy histories 
based on redshift zero observations through SED modeling is 
fraught with difficulty and ambiguity due to uncertainties in 
modeling star formation, stellar evolution, stellar structure, 
stellar atmospheres, chemical evolution, and dust
\citep{conroy10a, conroy13a, hayward15a, lower20a}. In this
paper we will provide simulations that will help us to 
better understand these uncertainties, especially those 
related to dust.

A particularly informative property of a galaxy is 
its  star formation rate (SFR), which tells us how 
rapidly the galaxy is converting gas into stars. Because young stellar populations emit much more strongly 
in the ultraviolet (UV) than do older populations, the 
UV luminosity emitted by stars is closely related to 
a galaxy's SFR. 

However, dust preferentially absorbs and scatters UV 
light relative to higher wavelengths and tends to be 
located near newly formed stellar populations \citep{salim20a}. 
Thus, the observed UV luminosity is typically substantially
less than the emitted luminosity. The combination 
of absorption and scattering due to a dusty medium between
the observer and a source is known as extinction,
and its wavelength dependence as the extinction curve.
The extinction curve depends on the distribution of dust
grain sizes and chemical composition.
The dusty medium and the light sources are 
intermixed, so in a real system some sources experience more 
extinction
than others. In addition, flux can be scattered by dust
into the line of sight.
In the presence of all of these effects, the ratio of the total observed 
to  emitted flux is known as the attenuation. The 
attenuation's dependence on wavelength can differ from 
the extinction curve of the dusty medium.

Correcting for the attenuation is a significant source of 
uncertainty in inferring the SFR of a galaxy. From galaxy
to galaxy, and within galaxies, there is variation in the 
amount of dust, in the dust extinction curves, and 
in the geometry of dust relative to the stars.
All of these differences can cause a substantial variation in 
the global attenuation curves.
Of course, we have no real way of knowing the true attenuation 
curve of a galaxy, since we don't have access to the emitted 
light before interacting with dust. This lack of a ground 
truth makes it impossible to test the accuracy of SED modeling 
inferences from observations alone.

In order to test our ability to infer physical properties of 
galaxies from observations, we need a data set for which ground
truth is known, and from which we can test our inferences. 
The approach we take here is to perform radiative transfer on 
simulated galaxies to produce mock photometry whose underlying 
physical parameters are known. 
This process has its  own difficulties, which stem from many of the 
same uncertainties  as in SED modeling, as well as difficulties related 
to the limited spatial and temporal resolution of the simulations.

Several recent investigations have pursued this approach.
\cite{camps16a} and \cite{trayford17a} describe 
methodology for generating mock photometry for EAGLE galaxies 
\citep{crain15a, schaye15a} and present comparisons with local 
galaxies from the Herschel Reference Survey \citep{boselli10a}.
They find discrepancies in the color-color relations in the 
$250$--$500 \mu m$ wavelength regime, indicating that the simulated 
dust is being insufficiently heated. The  EAGLE simulations 
\citep{schaye15a} do 
not model the cold gas phase of the interstellar medium (ISM), leading 
to an artificially smooth dust density structure which can result 
in less grey attenuation curves.  Related work by \cite{trcka20a} 
uses the same methodology described above with a more detailed 
comparison to observations from DustPedia and finds that their mock photometry has FUV and MIR luminosities that are consistently larger than observed galaxies at the same stellar mass. \cite{kapoor21a} generate mock photometry for 
30 galaxies from the Auriga project using a methodology similar to \cite{trayford17a}, but with some differences 
in the sub-grid treatment of star-forming regions. They also explore 
two different methods for assigning dust mass from the gas based 
on the temperature and density of the gas. Although they do compare 
some of their mock colors to observations, the presented comparisons 
are fairly limited in scope. The ISM in the Auriga galaxies is overly 
smooth on scales of a few hundred parsecs and does not 
reproduce the density structure of a multi-phase ISM \citep{marinacci19a}. 

As a step towards a more accurate accounting of the effects of dust 
when inferring star formation histories, we perform full radiative transfer 
using the Stellar Kinematics Including Radiative Transfer (SKIRT) 
\citep{camps20a} Monte Carlo radiative transfer code on simulated 
galaxies from the Numerical Investigation of Hundred Astrophysical Objects (NIHAO) project \citep{wang15a}, generating mock 
observations ranging from far-ultraviolet (FUV) to far-infrared (FIR) 
wavelengths. In contrast to the previous works described above, 
the simulated galaxies from the NIHAO suite do reproduce the density 
structure of a multi-phase ISM; 
however, the spatial smoothing  of
$\gtrsim 30$~pc is still
larger than the structures found in the observed ISM. 

We expand on previous works by presenting a detailed comparison between our mock observations and observed photometry from the nearby universe under a variety of modeling assumption, sub-grid recipes, and free parameter choices. The simulated photometry of our final modeling choices will be made publicly available as a data product and will provide an excellent basis on which to test the accuracy of SFHs inferred from SED modeling methods.

In Section \ref{Simulations} we describe the NIHAO simulations, and in Section 
\ref{Radiative Transfer} we describe our use of SKIRT for radiative transfer.
In Section \ref{Axis Ratios}, we describe the procedure for calculating axis ratios from simulated r-band images.
In Section \ref{Validating Against Observed Galaxies}, we discuss how we 
tune our parameters to reproduce the UV, optical, and infrared galaxy colors
and their axis ratio dependence as observed in DustPedia. 
In Section \ref{Results}, we present the  results for the galaxy colors  and
their axis ratio dependence, and the resulting attenuation curves. In Section \ref{Energy Balance}, we draw particular attention to the apparent energy imbalances that realistic
galaxy geometries can cause between UV dust attenuation and IR dust emission, highlighting the deficiency in the assumption of perfect energy balance commonly made in SED modeling codes such as Prospector \citep{leja17a}, MAGPHYS \citep{battisti20a}, and CIGALE \citep{boquien19a}.
We describe the data products available from this work in section \ref{Data Product}. We summarize our
results in Section \ref{Summary}. \\ 

\section{Simulations} \label{Simulations}

\noindent In this paper we use redshift zero snapshots from the NIHAO project, a set of more than one hundred cosmological zoom-in hydrodynamical simulations which reproduce the stellar versus halo mass relation, and the star formation rate versus stellar mass relation across cosmic time \citep{wang15a}. These simulations are based on the   $N$-body smoothed particle hydrodynamics (SPH) solver {\sc gasoline2} \citep{Wadsley2017}. 
The code includes cooling from hydrogen, helium, and metal-lines in a uniform UV ionizing background 
\citep[see][for more details]{shen10a,wang15a}, while feedback from supernovae
and massive stars (the so-called Early Stellar Feedback) is implemented following \cite{Stinson2013}.
NIHAO galaxies also include AGN feedback \citep{Blank2019} but for this study we limited ourselves to stellar feedback using the original sample of galaxies from \cite{wang15a}.

Gas particles are allowed to form stars when their temperature is below $15,000\,K$ and their density is above $10.3\,{\rm cm}^{-3}$. In the NIHAO simulations, each star particle represents a population of stars of the same age and metallicity (inherited from the parent gas particles) with a distribution of masses given by the assumed \citep{chabrier03a} initial mass function (IMF).

The NIHAO simulations have proven to be very successful in reproducing several observed
scaling relations like the relation between stellar mass and halo mass \citep{Wang2015}, the disk gas mass and disk size relation \citep{Maccio2016}, and the Tully-Fisher relation \citep{Dutton2017}, and have been recently quite successfully tested against a large variety of scaling relations based on the MaNGA sample \citep{arora2023}. \\

\section{Radiative Transfer} \label{Radiative Transfer}

\noindent SKIRT is a radiative transfer code specifically designed for dusty astrophysical systems like galaxies. It contains modules for importing data from smooth-particle hydrodynamic (SPH) simulations. In the following subsections, we motivate and describe our choice of wavelength grids, source models, dust models, and sub-grid recipes used in this work. We tune the free parameters within these models by comparing to the DustPedia sample \citep{davies17a}, which includes 875 galaxies in the nearby universe with over 40 aperture matched photometric bands ranging from the UV to millimeter wavelengths \citep{clark18a}, and which will be described in Section \ref{Validating Against Observed Galaxies}. \\

\subsection{Wavelength Grids} \label{Wavelength Grids}

\noindent To simulate the radiative transfer, we must choose the wavelength grids on which to describe various components of the light: the source radiation, the local radiation field, the dust emission, and the observed (``instrument'') SED. In order to maximize computational efficiency without losing resolution, we decouple the choices of wavelength grids for these components. The paragraphs below describes these choices; other than the increased optical resolution in the final set of simulations, these choices follow the recommendations of \cite{camps20a}.

For the source system, we set the wavelength grid to range from $0.04\, \mu {\rm m}$ to $2000 \, \mu {\rm m}$ with the same resolution as our source model, described in Section \ref{Source Model}. For the radiation field wavelength grid, which is used to heat the dust at every medium system cell, we use the range $0.04 \, \mu {\rm m}$ to $10 \, \mu {\rm m}$ with 40 bins distributed evenly in log space. For the dust emission wavelength grid, we use a nested logarithmic wavelength grid with a coarse component between $0.2 \, \mu {\rm m}$ and $2000 \, \mu {\rm m}$ with 308 bins, and a sub-grid component between $3 \, \mu {\rm m}$ and $25 \, \mu {\rm m}$ with 200 bins to give higher resolution in the region corresponding to polycyclic aromatic hydrocarbon (PAH) emission. 

\begin{figure}
\begin{center}
\epsscale{1.17}
\plottwo{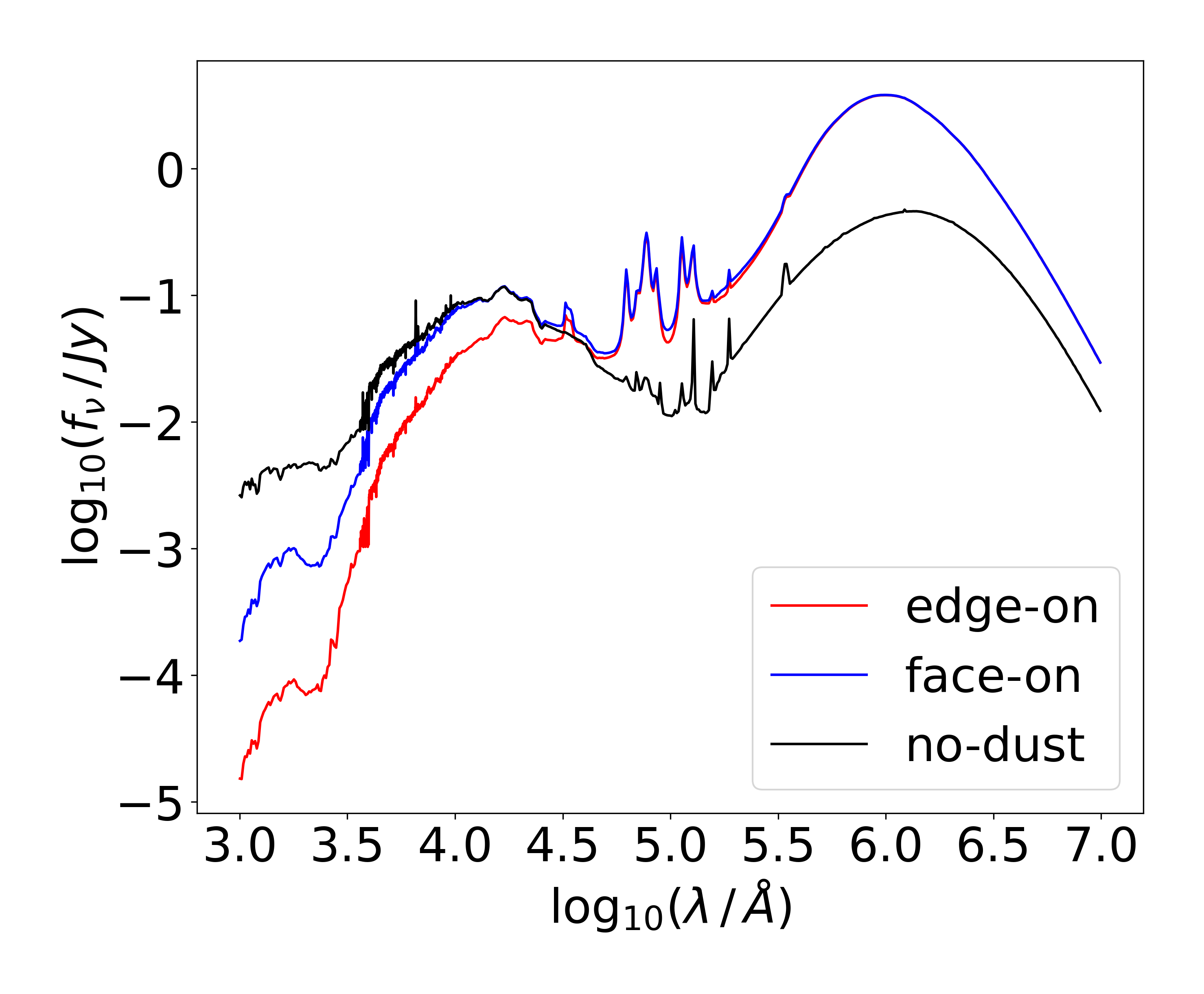}{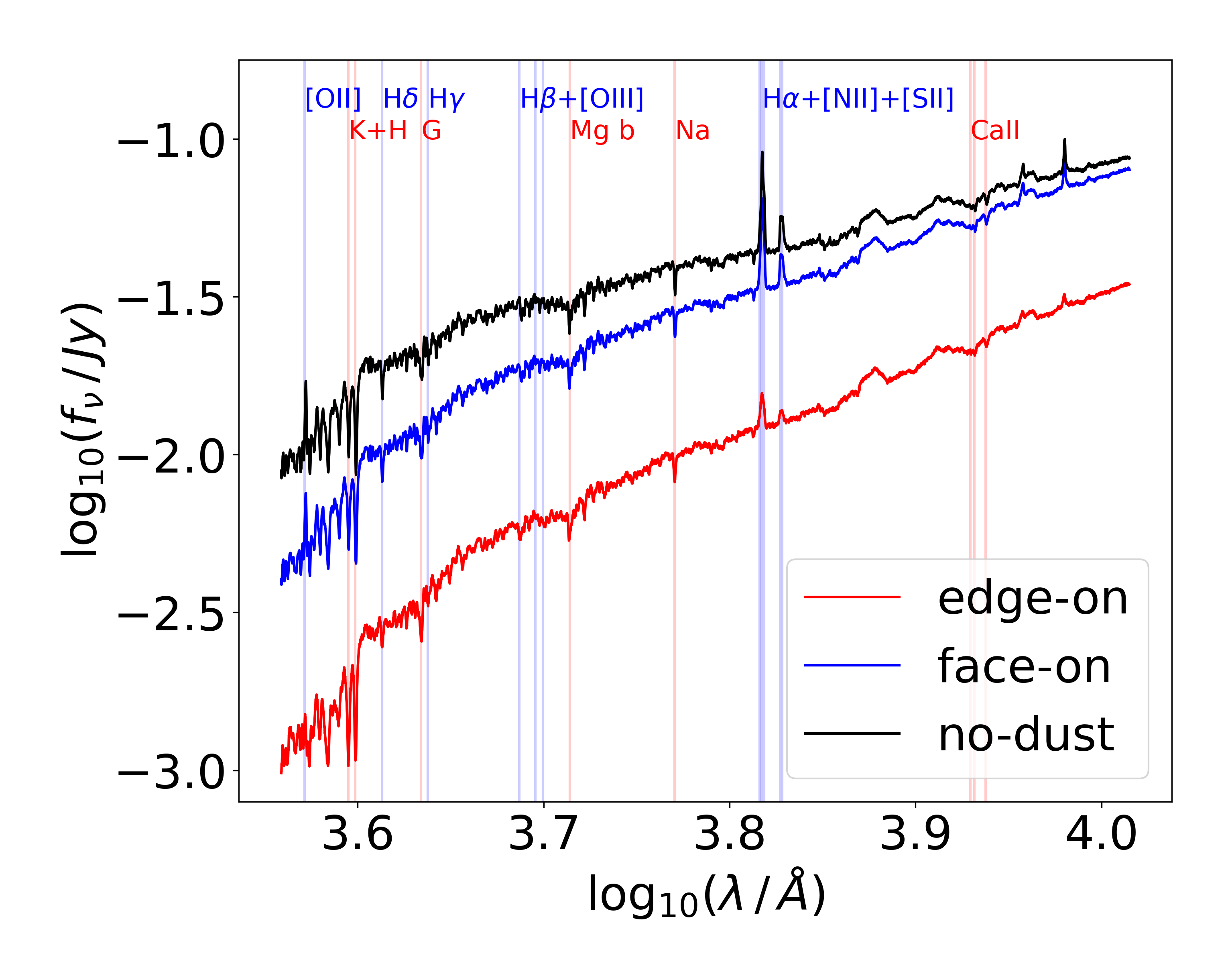}
\end{center}
\caption{\label{fig:sed} \small 
Spectral Energy Distributions (SEDs) of a NIHAO disk galaxy (g1.92e12) viewed from different orientations, and in the absence of dust, modeled at a distance of 100 Mpc with $\tau_{\rm clear}=2.5\,{\rm Myrs}$ and $f_{\rm dust}=0.1$. The plot on the left shows the SEDs from the far-ultraviolet to the far-infrared, the plot on the right shows the same SEDs zoomed in to the wavelengths corresponding to MaNGA with matched resolution. Vertical lines indicate emission (blue) and absorption (red) lines. Although the black SEDs are nominally ``dust-free,'' they still have a small amount of IR emission due to the \HII{} regions included in the spectra of the star-forming particles (even with $f_{\rm PDR}=0$); see section \ref{Source Model} for a quantitative analysis of this sub-grid attenuated energy. Note that the face-on orientation has less of its UV-optical light attenuated while having nearly identical IR emission. This comparison alone shows that strictly following energy balance when modeling dust attenuation and emission along a particular line of sight is not well motivated.}
\end{figure}

For our SED instrument wavelength grid, we use two different choices, one for tuning parameters and one to produce the final set of simulations presented here. For tuning parameters, we use 601 bins evenly distributed in log space between $0.04 \, \mu {\rm m}$ and $2000 \, \mu {\rm m}$. For producing the final presented set
of simulations, we use the same grid except that we
increase the SED instrument wavelength resolution between 3600 and 10,300 
\AA\ to a resolution of $R \sim 2000$ to match optical spectra from 
the Sloan Digital Sky Survey IV's Mapping Nearby Galaxies at APO program (SDSS-IV MaNGA, \citealt{bundy15a, blanton17a}). See Figure \ref{fig:sed} for visualization of the instrument wavelength range and resolution.

\subsection{Source Model} \label{Source Model}

\noindent For each star particle, we assign a spectral energy distribution (SED) from FSPS \citep{conroy09a, conroy10a, foreman-mackey14a} according to its age and metallicity, assuming a Chabrier initial mass function (IMF) \citep{chabrier03a}, and weighted by its mass. We sample photons from the SED and launch them isotropically in the rest frame of the particle and subsequently Doppler shift them accordingly. 

In the real universe, a cloud of cold gas will collapse gravitationally to form a population of stars. These populations of stars typically have total stellar masses between $700$ and $10^6 M_{\odot}$, distributed according to a power law with index $-1.8$ \citep{heyer01a}. In contrast, the populations of stars in the simulations are represented by single-age, single metallicity star particles which typically have much larger masses. This approximation leads to star formation histories that are more stochastic than they would be if simulated at higher resolution. For old stellar populations, this effect is insignificant, but for young stellar populations, whose spectra change rapidly with slight changes in age, this effect will cause an unrealistically large variation in the resulting spectra. 

\begin{figure}
\begin{center}
\epsscale{1.17}
\plottwo{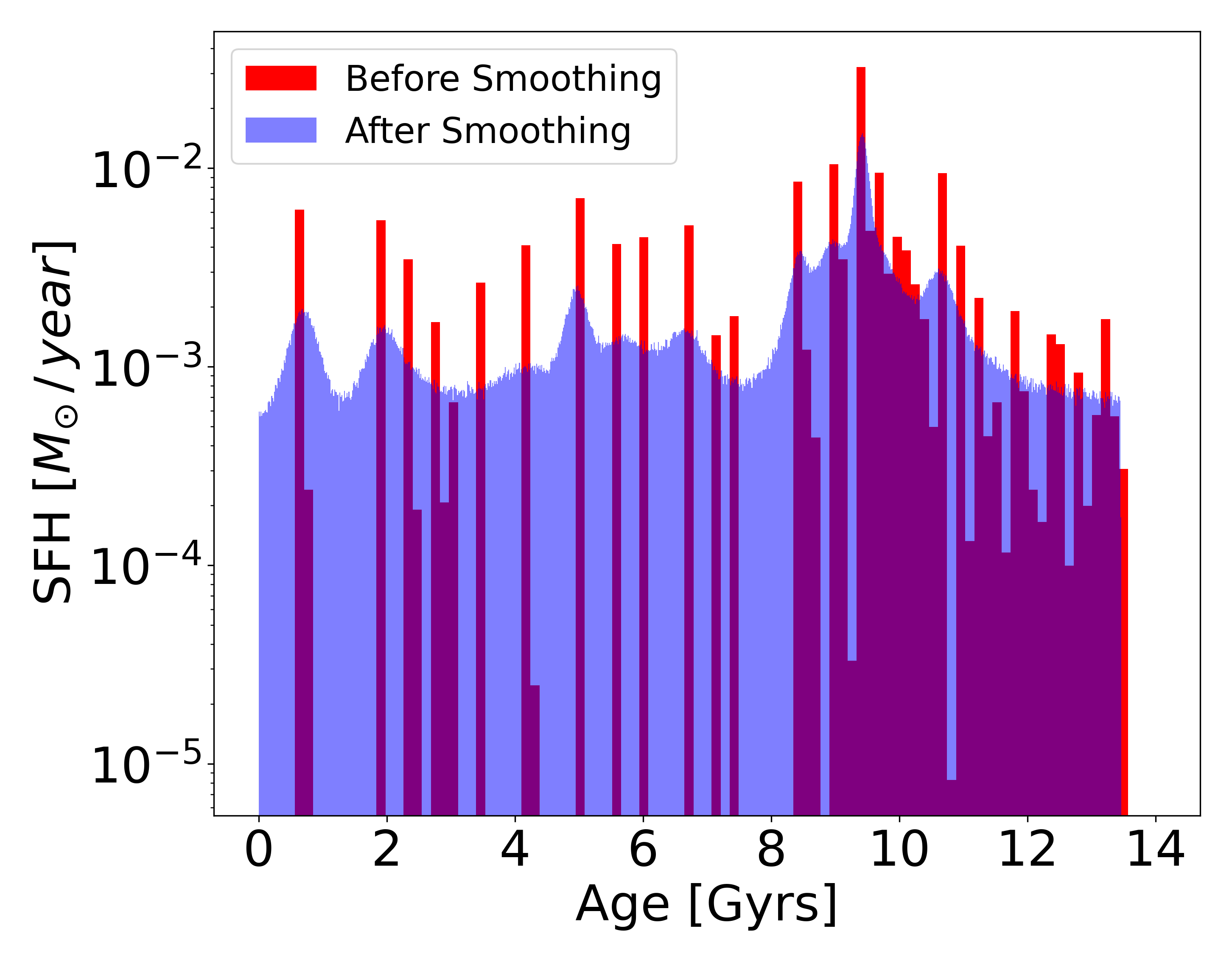}{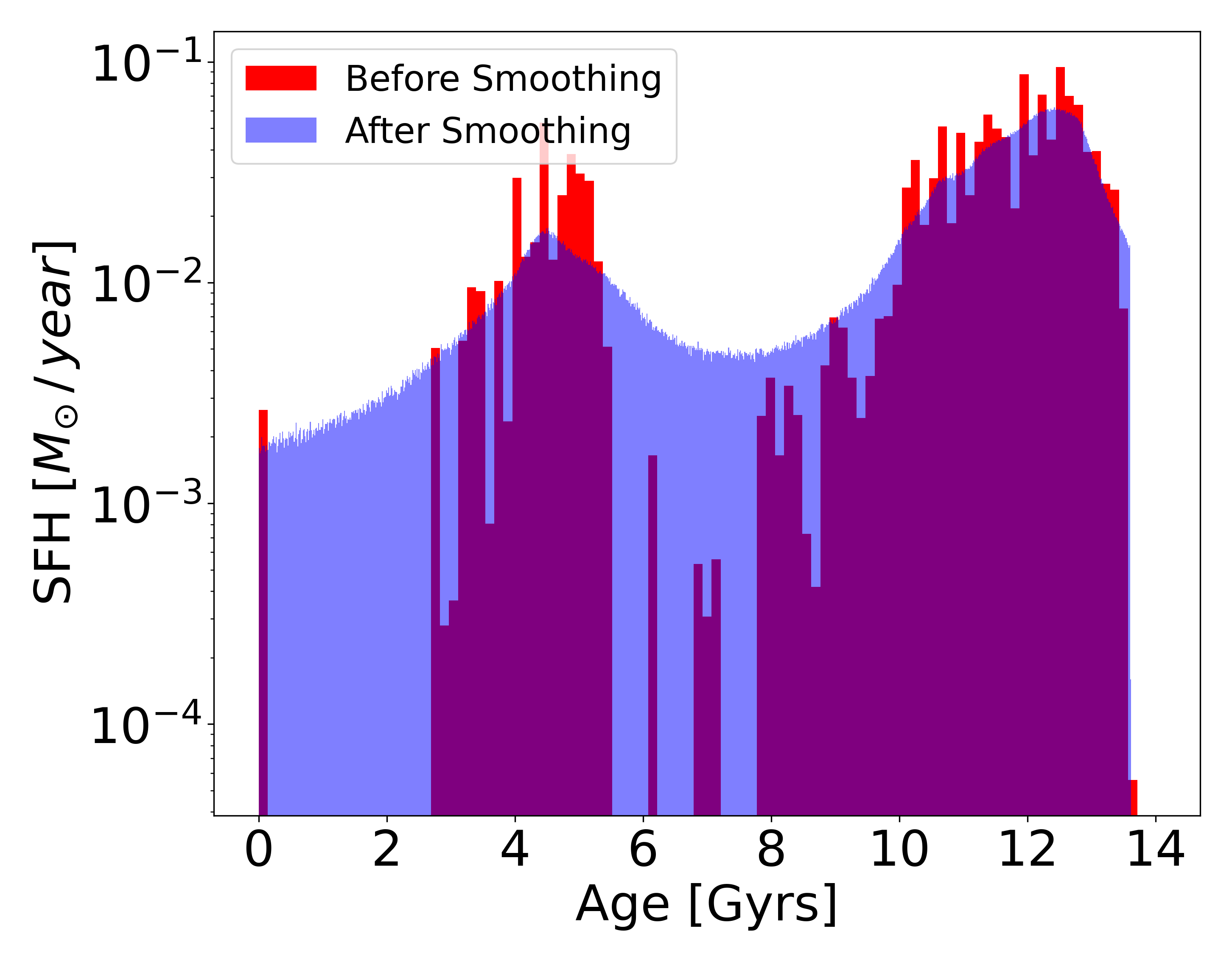}
\end{center}
\caption{\label{fig:SFHs} \small 
SFHs before and after smoothing for two different galaxies. SFH after smoothing is binned 10 times more finely.}
\end{figure}

\begin{figure}
\begin{center}
\includegraphics[width=0.65\textwidth]{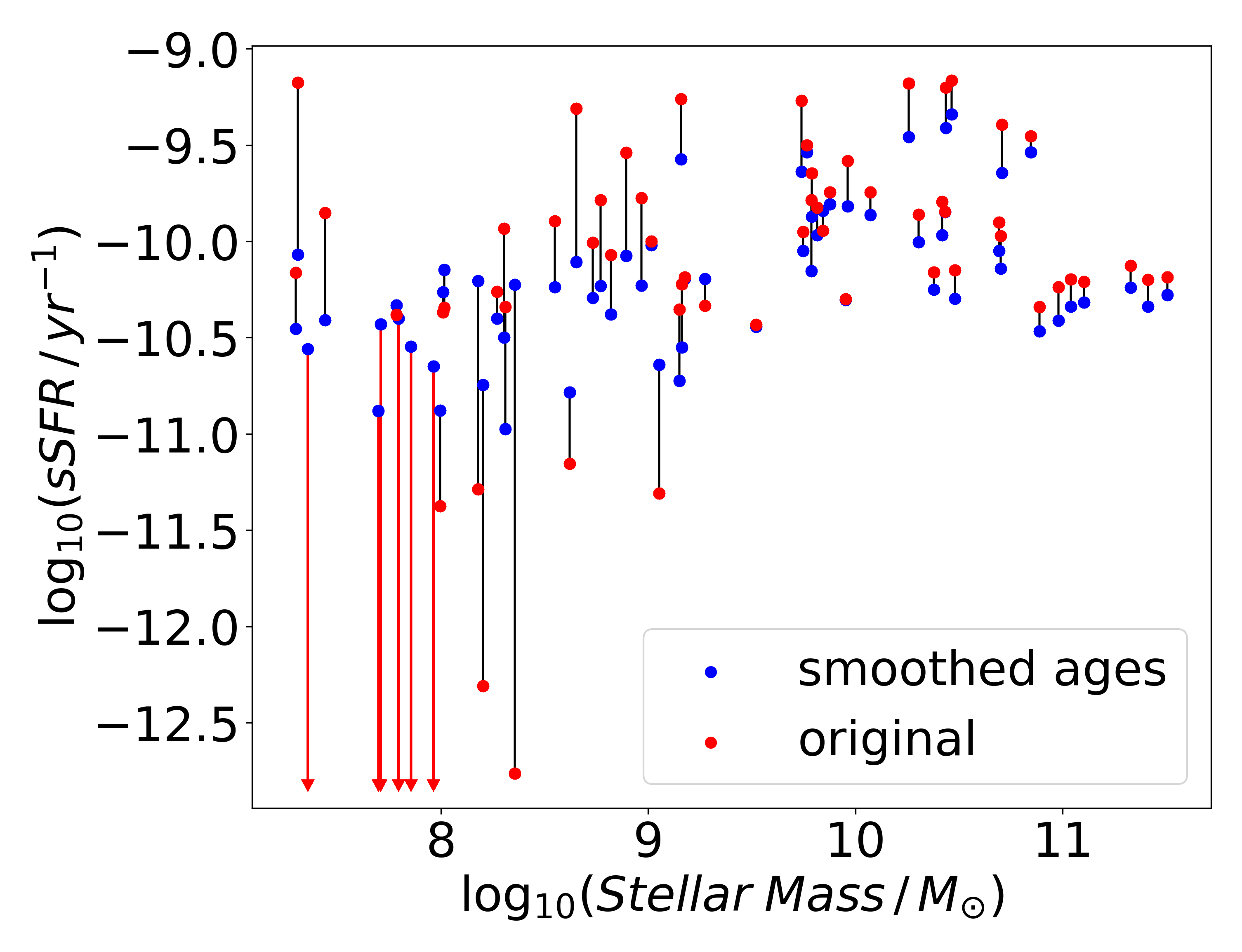}
\end{center}
\caption{\label{fig:ageSmooth_sSFR_stellarMass} \small 
Distribution of NIHAO galaxies' sSFR and total stellar mass before and after age smoothing. Red arrows indicate an original sSFR equal to zero.}
\end{figure}

\begin{figure}
\begin{center}
\includegraphics[width=0.88\textwidth]{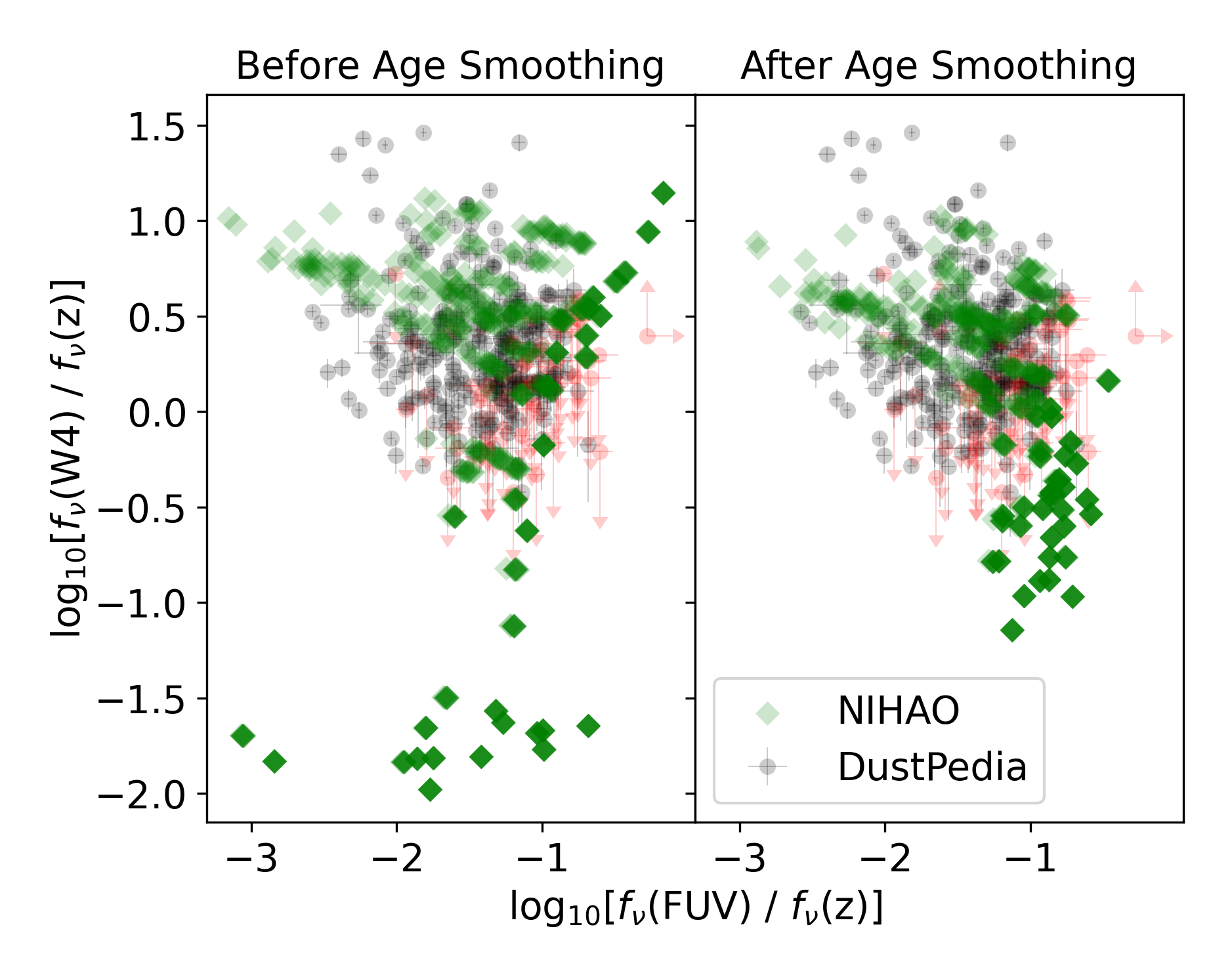}
\end{center}
\caption{\label{fig:ageSmoothColorColorPlots} \small 
$\log_{10}[f_{\nu}({\rm W4})/f_{\nu}(z)]$ vs. $\log_{10}[f_{\nu}({\rm FUV})/f_{\nu}(z)]$ before and after the applications
of the age smoothing recipe with $\tau_{\rm clear}=2.5\,$Myrs 
and $f_{\rm dust}=0.1$.}
\end{figure}

To mitigate this effect, we have developed and implemented a sub-grid recipe which effectively smooths out the simulated star-formation histories such that the typical difference in age between two neighboring (in the temporal sense) young star particles is less than $\sim 1$ Myr, the timescale on which these spectra show significant variation. To accomplish this, we replace each star particle with 50 new particles of equal mass (normalized to conserve total stellar mass) and with ages sampled from a normal distribution centered around the age of the parent particle. The standard deviation of the distribution is calculated as the larger of the two age differences between the parent particle and the $k^{\rm th}$ younger and older neighbors, multiplied by a constant scaling parameter, $S$. We also ensure that sampled particles have ages greater than $0$ and less than the age of the oldest parent particle by re-sampling any particles which do not meet these criteria. Through visual inspection of the smoothed SFHs, we found that $k=100$ and $S=100$ produce sufficiently high resolution distributions of young star particles while still retaining the characteristic features of the original SFHs (see Figure \ref{fig:SFHs} for comparisons). The effect of this age smoothing procedure on the sSFR averaged over the last 100 Myrs is shown in Figure \ref{fig:ageSmooth_sSFR_stellarMass}. We can see that most significant changes in sSFR occur in the lower mass galaxies. 
This is because these galaxies have fewer star particles, so when we only count 
star particles created in the last 100 Myrs, the scatter in the sSFR is 
dominated by the discreteness of the particles rather than physical
effects within the simulation. Looking at Figure \ref{fig:ageSmoothColorColorPlots}, we can see that the age smoothing procedure brings sSFR sensitive colors closer to the distribution of colors found in DustPedia. 

For star particles younger than 10 Myrs, we also need to account for the absorption and emission by dust within photo-disassociation regions (PDRs) that result from the remaining birth clouds of newly formed stars. Because these regions are below the spatial resolution of the simulations, we adopt the commonly used method \citep{groves08a, jonsson10a, hayward15a, trayford17a, trcka20a, kapoor21a} of assigning SEDs from MAPPINGS-III that already include the effects of photoionization and obscuration within these dense molecular clouds. The underlying stellar SEDs used in these calculations are from Starburst99 \citep{leitherer99a}. 
The emergent SED spectra are characterized by age, metallicity, ISM pressure, and the PDR covering fraction. Age and metallicity are properties of the star 
particles in NIHAO. Following \cite{kapoor21a}, we stochastically sample compactness parameters, a measure of the density of \HII{} regions, from the log-normal distribution with $\langle$log C$\rangle$ = 5 and a standard deviation of 0.4. We then calculate ISM pressures for each sampled compactness parameter using Equation 13 from \cite{groves08a}. The PDR covering fraction is calculated for each particle using the equation $f_{\rm PDR} = \exp(-t/\tau_{\rm clear})$ where $t$ is the age of the particle and $\tau_{\rm clear}$ is a free parameter of our model. We also introduce ``ghost" gas particles with negative mass equal to 10 times the stellar mass of the parent particle to avoid double counting the dust in these regions, following \cite{trayford17a}.

We will produce ``dust-free'' spectra by excluding the interstellar dust associated
with simulation gas particles described in the next subsection. We produce 
these spectra in order to calculate attenuation curves. However, 
as noted in Figure \ref{fig:sed}, in the \cite{groves08a} models for young
stellar particles there is some dust attenuation and IR emission, even when $f_{\rm PDR}=0$. 
Because we only use the dust-free spectra to calculate attenuation, as
long as the attenuation internal to the young stellar particles is small
enough we can safely ignore this effect.
In order to quantify this  attenuated energy, we calculate the energy between 
912 \AA\  and 2 microns for FSPS particles (which are truly dust-free) with 
100 ages evenly spaced between 0 and 10  Myrs, and one MAPPINGS-III particle (which assumes a constant SFR over the past 10 Myrs), both normalized to 
$1 \, M_{\odot}$. We use FSPS here instead of Starburst99, which is the underlying SSP used in the MAPPINGS-III star-forming particles, because the version of Starburst99 used in SKIRT also includes IR emission and therefore must also have some of its energy attenuated already. From these calculations, we find that 
3.41\% of the energy in the $f_{\rm PDR}=0$ MAPPINGS-III particle spectra has been attenuated relative 
to FSPS. Furthermore, since only a small fraction of the stellar mass is 
contained in these star-forming particles, we conclude that this 
sub-grid attenuation in the ``dust-free'' spectra is insignificant to the resulting attenuation curves. \\ \\ 

\subsection{Dust Model} \label{Dust Model}

\noindent Because the NIHAO simulations do not directly model the dust population, we assume that each gas particle contains a dust mass given by a fixed fraction, the dust-to-metal ratio, of its mass in metals. We constrain the dust-to-metal ratio, the second free parameter of our model, by comparing the resulting photometry to the DustPedia sample. We also assume that no dust is present in gas above a maximum temperature of $16,000 \, K$ although the results presented in this paper are insensitive to changes of $\sim 10 \%$ to this temperature. 

To perform radiative transfer calculations, we discretize space using an OctTree spatial grid which we subdivide until each grid cell contains at most one gas particle. This discretization procedure is overridden by parameters which set the minimum and maximum allowed refinement levels. We set the minimum level to 6 and the maximum to 99, the largest allowed value, so that we don't override the one gas particle per cell discretization procedure. Although lowering the minimum level has only a small effect on the global photometry, we choose this value for the purpose of generating higher resolution images in the diffuse ISM. Once the spatial grid has been set, each cell is considered to be uniform in its optical properties.

We utilize the THEMIS dust model \citep{jones17a} in our radiative transfer calculations. Of particular importance to this work, this dust model attributes different extinction curves depending on the density in each cell of gas, reflecting the evolution of dust in response to the physical conditions of the ISM, with more aggregated grains residing in denser regions. Rather than being tuned to fit astronomical observations, the dust optical properties in this model are determined from laboratory measurements of cosmic dust analogues. This model considers a two component dust population consisting of amorphous silicates and amorphous hydrocarbons. The silicate grains, which are half amorphous enstatite and half amorphous forsterite by mass, are assumed to follow a log-normal size distribution while the hydrocarbon grains are assumed to follow a combination of log-normal and power-law distributions. We use 15 size bins for both the silicate and the hydrocarbon populations, each of which is heated stochastically by the local radiation field as described in \cite{camps15a}. Relative to assuming local thermal equilibrium, the stochastic heating of grains has the effect of widening the distribution of dust temperatures since smaller grains will have higher temperatures and larger grains will have lower temperatures for a given radiation field. \\

\section{Axis Ratios} \label{Axis Ratios}
\noindent In order to find viewing orientations that span the range of each galaxy's axis ratios in an efficient way, we run separate radiative transfer simulations that only include star particles (no dust) and only store the $r$-band fluxes, denoted here as $f_r$, spatially resolved in 250$\times$250 pixels. For each simulated galaxy, the SKIRT parameter file is populated with 500 instruments with inclination and azimuth angles sampled uniformly on a sphere (the roll angle rotates along the line of sight and thus doesn't affect axis ratios).

To calculate axis ratios for each orientation, we used a method based on 
flux-weighted moments of the dust free image. Since galaxies do not have 
purely elliptical profiles, the axis ratio is not a perfectly well-defined 
quantity. We are interested in the axis ratio of the disk, not the bar or
bulge or the stellar halo, and have adjusted the flux weighting technique 
to be most sensitive to the disk region.

We first find the center of each image by taking the average position 
of the 50 brightest pixels. We then use the value of the $50^{th}$ brightest 
pixel as the maximum allowed flux value; all larger fluxes are set to 
this value so that they don't dominate the fits. After normalizing the 
fluxes such that the maximum value is unity, we set all pixels with normalized fluxes value less than 0.01 to 0 to avoid 
fitting to the galactic outskirts. Next we scale the normalized pixel 
fluxes using $\sinh^{-1}(f_r/0.01)$ so that the fits are sensitive 
to the light coming from disks, not only the brightest regions. Finally, 
we mask out pixels whose distance is less than 7 pixels away from the 
center ($r < 7$) to keep pixels with small radii from dominating the fits. 

Then we calculate the second moments of the adjusted fluxes, converting
them to position angles and axis ratios using the method described by
\cite{lupton1996a}. The second moments are:
\begin{eqnarray} \label{eq:Ms}
M_{xx} &=& \frac{\sum{(f_r \, x^2) / r^2}}{\sum{f_r}} \cr
M_{yy} &=& \frac{\sum{(f_r \, y^2) / r^2}}{\sum{f_r}} \cr
M_{xy} &=& \frac{\sum{(f_r \, x \, y) / r^2}}{\sum{f_r}}
\end{eqnarray}
In analogy to Stokes' parameters for polarization, we define:
\begin{eqnarray} \label{eq:QU}
Q &=& 2 M_{xx} - 1 \cr
U &=& 2 M_{xy}
\end{eqnarray}
Then the position angle and axis ratio can be written:
\begin{eqnarray} \label{eq:phiba}
\phi &=& \frac{1}{2} \tan^{-1}\left(\frac{U}{Q}\right) \left(\frac{180^\circ}{\pi}\right) \cr
\frac{b}{a} &=& \frac{1 + Q^2 + U^2 - 2 \sqrt{Q^2 + U^2}}{1 - \left(Q^2 + U^2\right)}
\end{eqnarray}

The result of this analysis is a table of inclination and azimuth 
angles and their corresponding axis ratios. From this table we select 10 orientations 
for each galaxy which span its range of axis ratios as evenly as possible. \\

\section{Validating Against Observed Galaxies} \label{Validating Against Observed Galaxies}

\begin{figure}
\begin{center}
\includegraphics[width=0.7\textwidth]{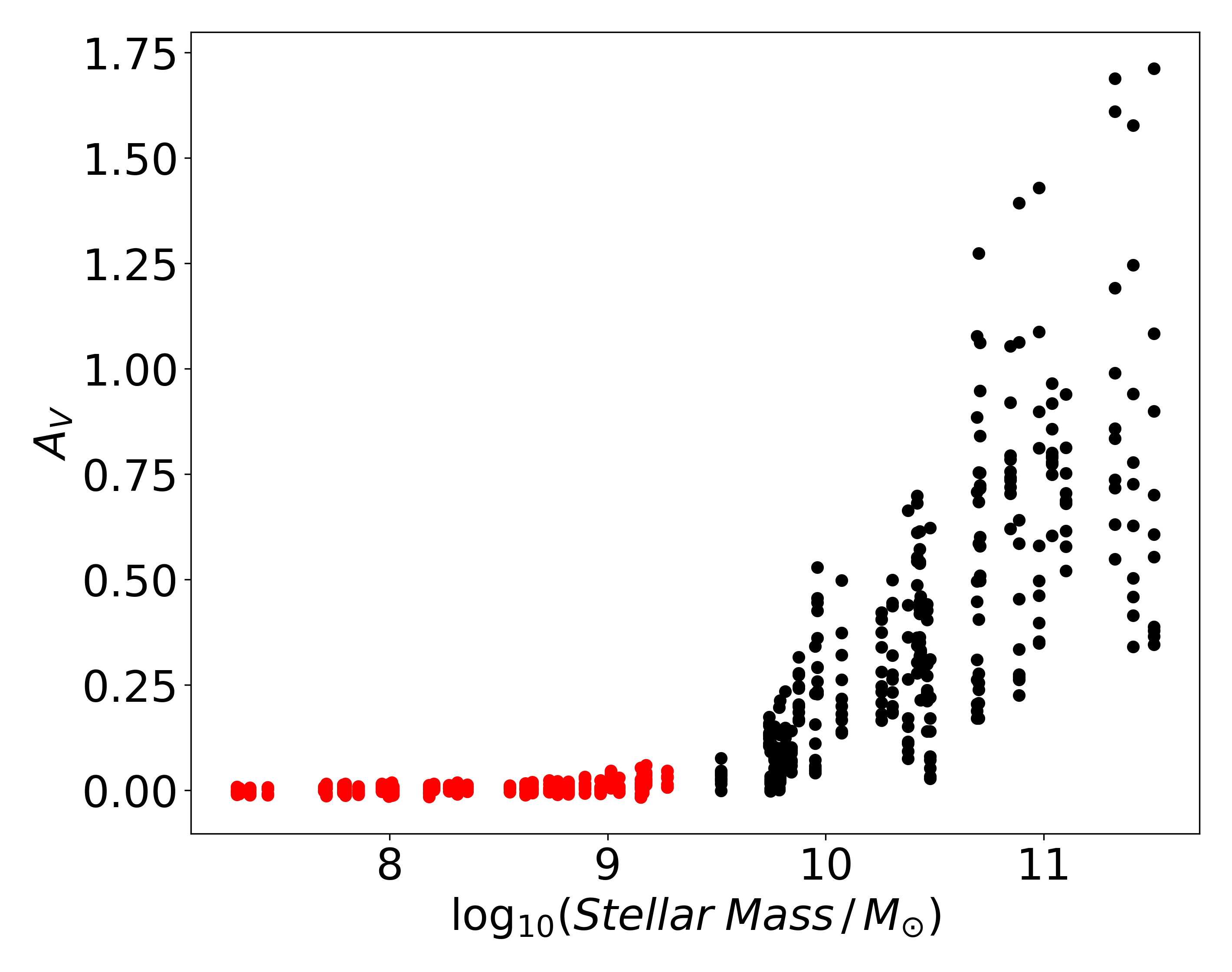}
\end{center}
\caption{\label{fig:AvStellarMass} \small 
$A_V$ vs. Stellar Mass for all 10 selected orientations of each NIHAO galaxy with $\tau_{\rm clear}=2.5\,$Myrs and $f_{\rm dust}=0.1$. It is clear that very small values of $A_V$ are strongly correlated with low stellar mass galaxies. Galaxies with total stellar mass below $10^{9.5}\,M_{\odot}$ are plotted in red.}
\end{figure}

\begin{figure}
\begin{center}
\epsscale{1.17}
\includegraphics[width=0.88\textwidth]{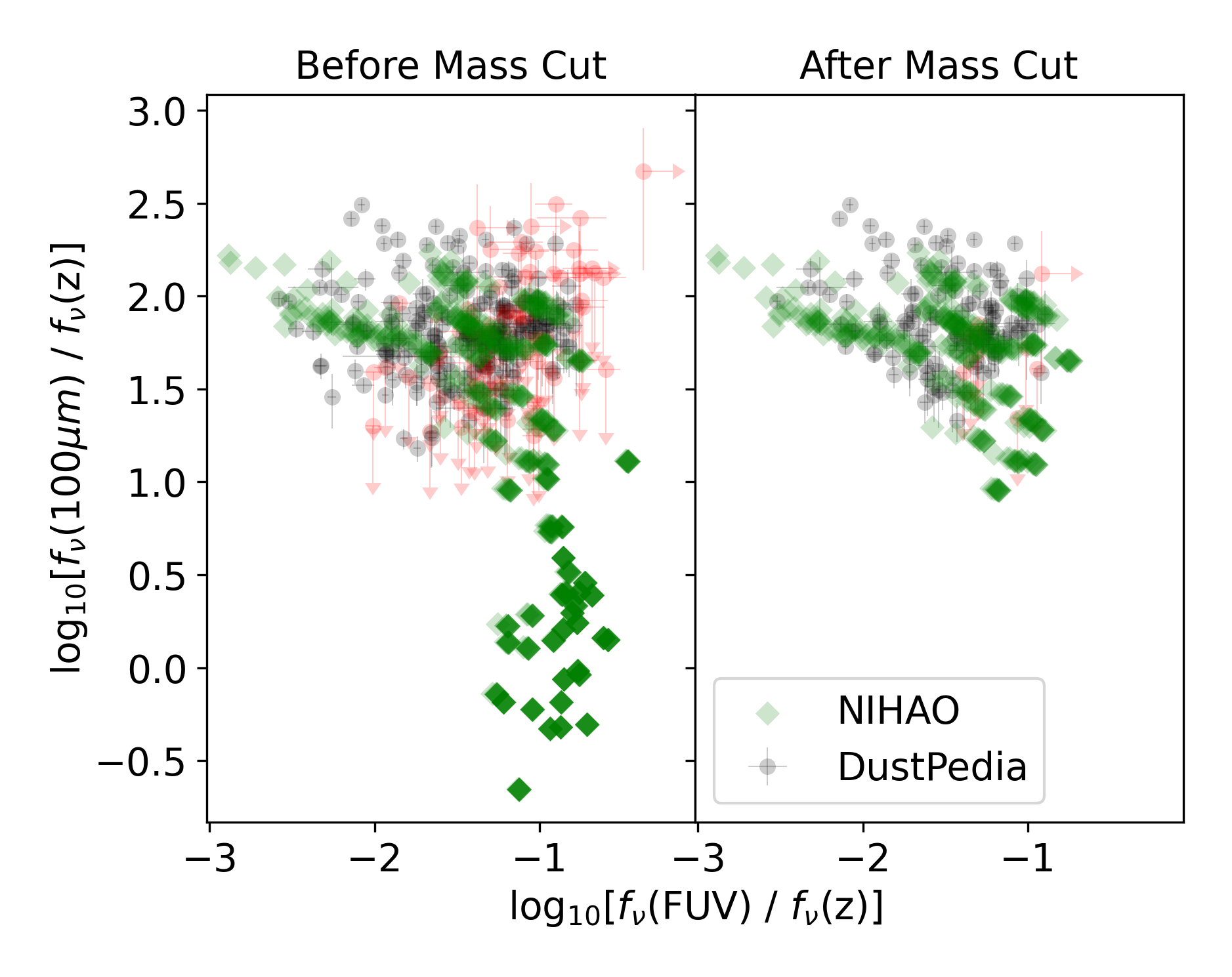}
\end{center}
\caption{\label{fig:colors_massCut} \small 
Flux Ratios $\log_{10}[f_\nu(100\mu m)/f_\nu(z)]$ vs. $\log_{10}[f_\nu(\rm FUV)/f_\nu(z)]$ before (left) and after (right) excluding NIHAO galaxies with total stellar mass below $10^{9.5}\,M_{\odot}$ with $\tau_{\rm clear}=2.5\,$Myrs and $f_{\rm dust}=0.1$. The red points are DustPedia galaxies that can only provide a limit on the flux ratio. In each plot, we only show DustPedia galaxies with stellar masses and sSFRs which are within the corresponding NIHAO ranges.}
\end{figure}

\begin{figure}
\begin{center}
\epsscale{1.17}
\includegraphics[width=0.8\textwidth]{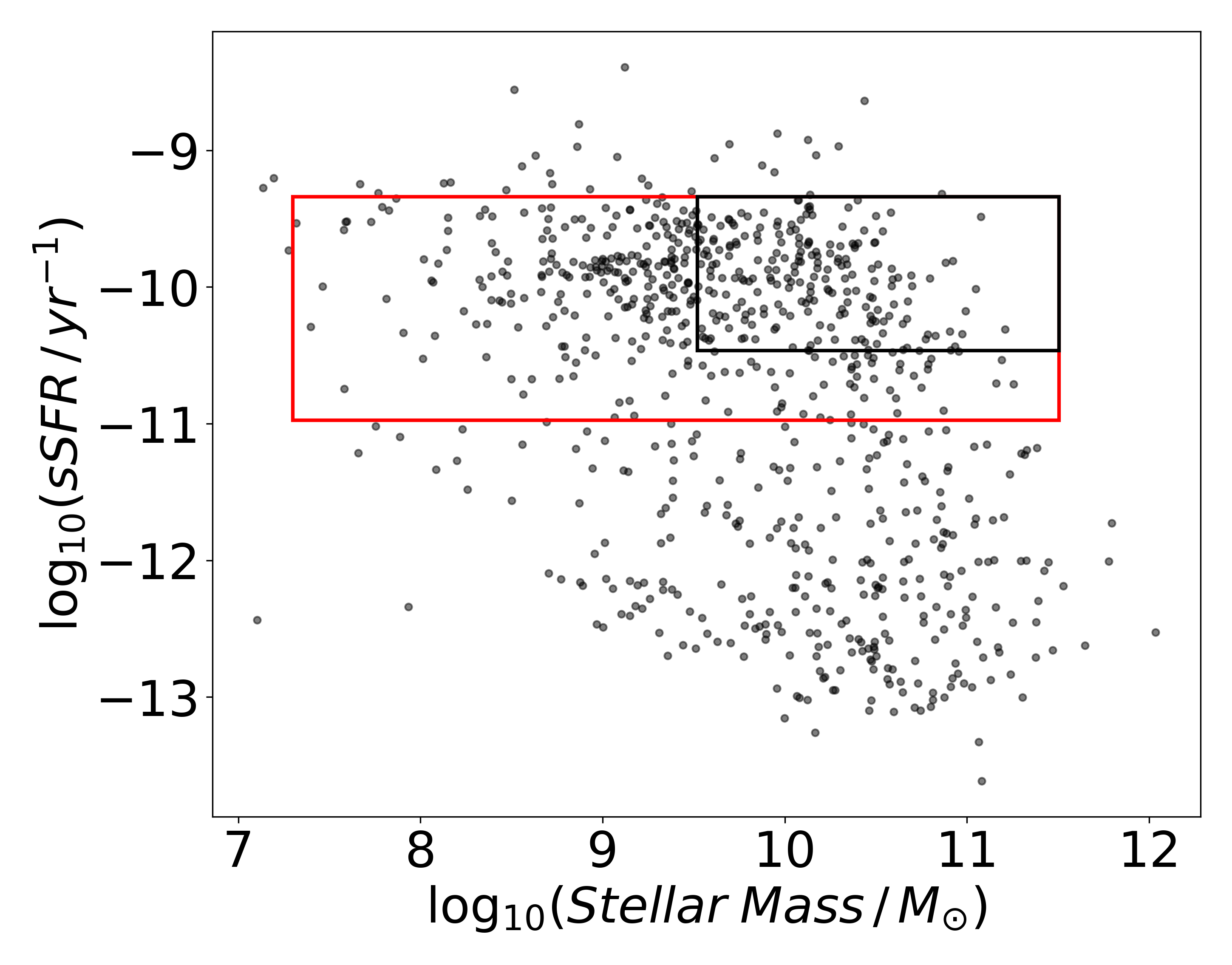}
\caption{\label{fig:dp_mass_sSFR} Distribution of sSFR vs. Stellar Mass for the DustPedia sample. The inlaid boxes show the sSFR and stellar mass ranges of the NIHAO galaxies before (red) and after (black) excluding galaxies with stellar mass below $10^{9.5}\,M_{\odot}$.}
\end{center}
\end{figure}

\noindent As mentioned in Section \ref{Source Model}, we use the DustPedia sample \citep{davies17a} to validate our radiative transfer parameter choices. This sample includes mass and SFR estimates for each galaxy as inferred from CIGALE fits \citep{boquien19a}. For the purpose of comparing these galaxies with the mock observations from the NIHAO sample, we exclude any simulated galaxy with a total stellar mass below $10^{7} \, M_{\odot}$. For all photometric comparisons presented in this work, we only include DustPedia galaxies with total stellar masses and sSFRs which fall within the range of the NIHAO galaxies being considered. 

For the purpose of constraining our free model parameters, we only want to include DustPedia galaxies that are reliable in their FUV and IR photometry. Figure \ref{fig:AvStellarMass} shows the stellar masses and $A_V$ values,
the difference between the attenuated and unattenuated V-band 
($\lambda \sim 5500\,$\AA) magnitudes, 
for all chosen orientations for all galaxies in our simulations. We can see a strong correlation between $A_{V}$ and stellar mass.
In particular, we see that galaxies with stellar mass below $\sim 10^{9.5}\,M_{\odot}$ experience very little attenuation regardless of axis ratio. These low $A_V$ value galaxies have a strong affect on dust-sensitive color-color plots. Figure \ref{fig:colors_massCut} shows the distributions of $\log_{10}[f_\nu(100\mu 
m)/f_\nu(z)]$ vs. $\log_{10}[f_{\nu}(\rm FUV)/f_{\nu}(z)]$ colors for DustPedia and NIHAO galaxies. The left panel shows both full samples, and
the right panel shows the samples
limited to galaxies with total stellar mass above $10^{9.5}\,M_{\odot}$. 
Red points represent DustPedia galaxies with one or more photometric fluxes 
that are consistent with zero at 2$\sigma$, and are plotted as upper or lower 
limits. Note that each plot only contains DustPedia galaxies that fall 
within the sSFR and stellar mass ranges of the corresponding NIHAO set. We can see from this figure that DustPedia galaxies which fall within the total stellar mass and sSFR ranges of the low mass NIHAO galaxies, but not the higher mass NIHAO galaxies, tend to only give upper limits on the $\log_{10}[f_\nu(100\mu m)/f_\nu(z)]$ colors. Because of this, these low mass NIHAO galaxies will not be useful in constraining our free model parameters. For the rest of this paper, unless otherwise stated, we will exclude NIHAO galaxies with $M_\ast < 10^{9.5}$ $M_\odot$ from our analysis. See Figure \ref{fig:dp_mass_sSFR} to see the distribution of inferred total stellar masses and sSFRs of the DustPedia galaxies, along with the NIHAO ranges before and after the exclusion of these low mass galaxies.

Given the radiative transfer model choices described above, there are two free parameters that need to be constrained by comparison to observations. The
first is the dust-to-metal ratio, which acts as a scaling parameter for the amount of 
diffuse dust in the interstellar medium. The second is the birth cloud clearing
time, which determines the covering fraction of sub-grid dust assumed to exist
around the very youngest  stellar populations as a function of their age. 

By comparing color-color plots that are sensitive to the effects of dust between the DustPedia sample and our mock observations 
of the NIHAO galaxies, we can determine whether or not our parameter 
choices are producing realistic photometry. Whereas previous works 
\citep{camps16a, trayford17a, trcka20a, kapoor21a} have tuned parameters to agree with observed color-color distributions, 
here we consider an additional constraint related to how galaxies' 
colors depend on viewing angle (the effects measured by
\citealt{maller09a}, \citealt{masters10a}, and \citealt{conroy10c}). In particular, we perform separate linear 
fits to the dependence of $\log_{10}[ f_\nu({\rm FUV})/f_\nu(z)]$ and 
$\log_{10}[ f_\nu(i)/f_\nu(z)]$ flux ratios on axis ratio 
for disk galaxies in NIHAO and DustPedia. The difference in slope 
of these fits between simulations and observations acts as a metric for 
determining the accuracy of our post-processing parameter space. 

\clearpage

\section{Results} \label{Results}

\noindent In Section \ref{Axis Ratio Color Plots} we quantify how galaxy colors vary with axis ratio. In Section \ref{Parameter Space} we search the parameter space of our radiative transfer model and determine the parameters that best match observations. Once we have finalized our parameter choices, we present simulated galaxy images, 
color distributions, attenuation curves, and energy balance as a function of axis ratio (Sections \ref{Images}, 
\ref{Color-color Plots},  \ref{Attenuation Curves},
and \ref{Energy Balance}).

\subsection{Axis Ratio Color Plots} \label{Axis Ratio Color Plots}

\noindent In this section, we determine the dependence of 
$\log_{10}[f_\nu({\rm FUV})/f_\nu(J)]$  and $\log_{10}[f_\nu(i)/f_\nu(J)]$ 
flux ratios on axis ratio for DustPedia disk galaxies and 
NIHAO disk galaxies. For NIHAO galaxies, we define a disk galaxy as one whose 
minimum axis ratio across the sampled orientations is less than $0.2$. For 
DustPedia, we use the morphological classifications included with the data and 
consider any numerical Hubble stage which is greater than or equal to 0 to be 
a disk galaxy. Qualitatively, we expect 
$\log_{10}[f_\nu({\rm FUV})/f_\nu(J)]$  and $\log_{10}[f_\nu(i)/f_\nu(J)]$ 
colors to increase with increasing axis ratio, because the light emitted 
from stars will on average need to travel through more dust as our line of 
sight becomes more edge-on. 

\begin{figure}
\begin{center}
\epsscale{1.17}
\plottwo{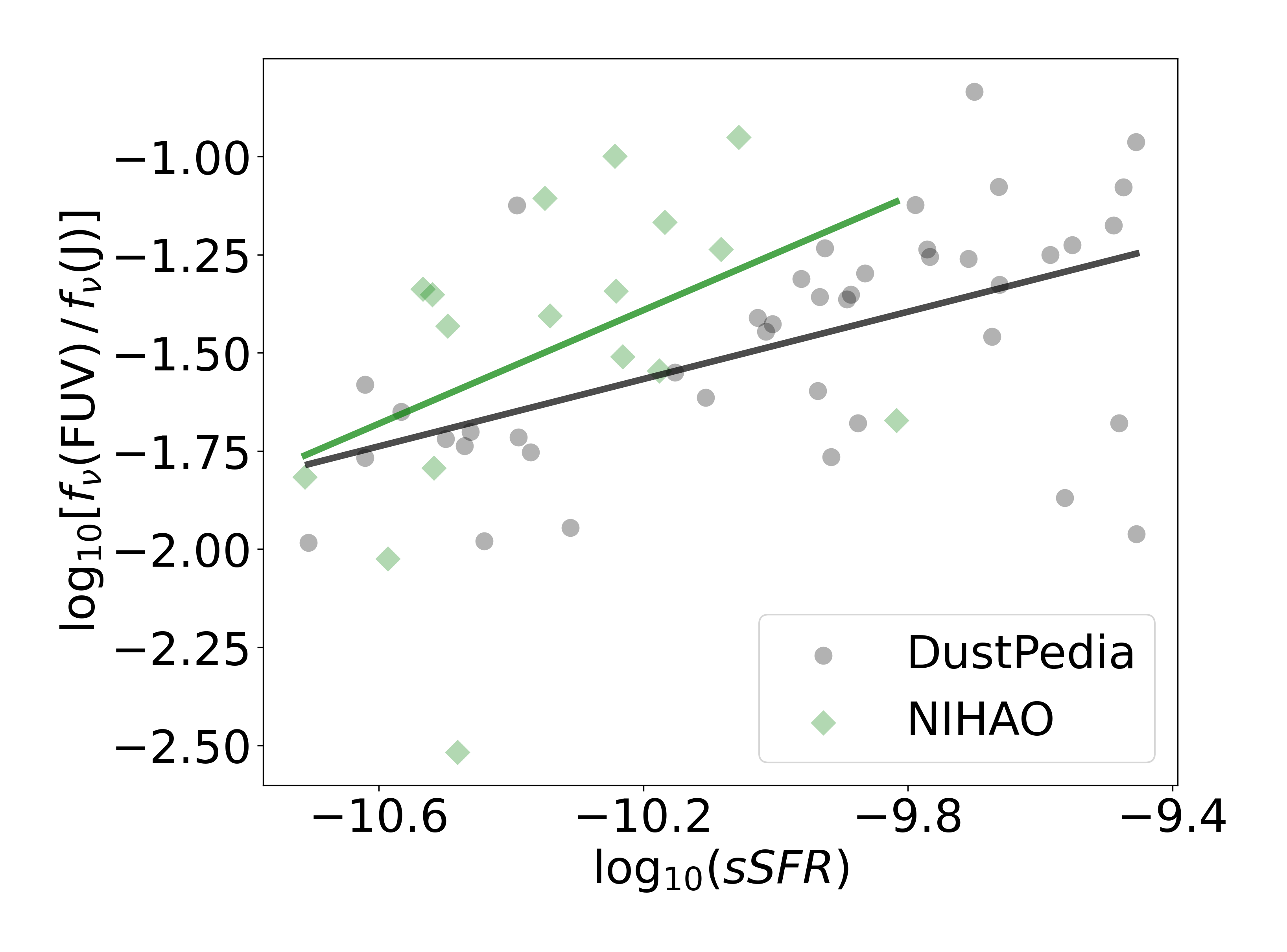}{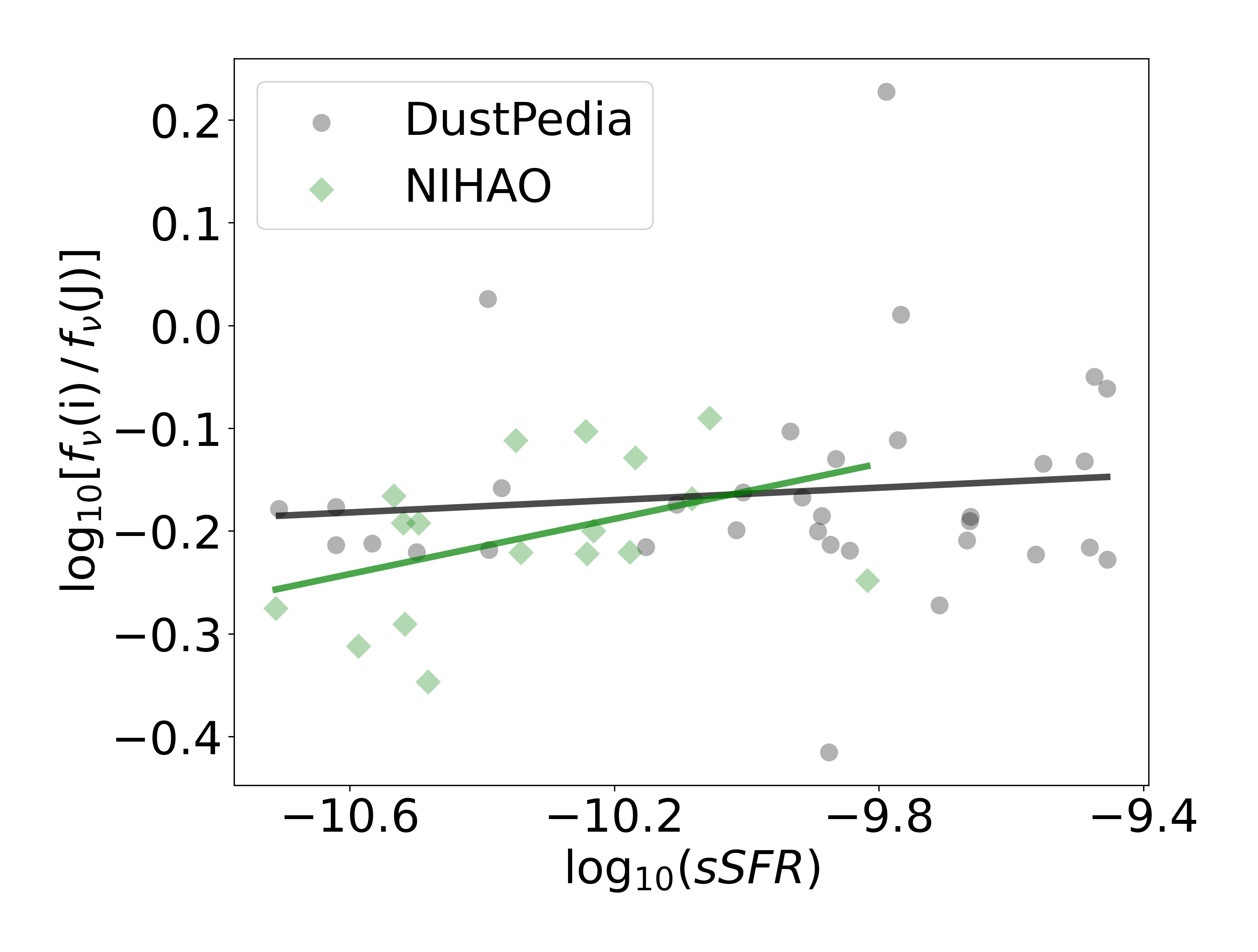}
\end{center}
\caption{\label{fig:FitPlots} \small 
Flux ratios $\log_{10}[f_\nu(\rm FUV)/f_\nu(J)]$ and $\log_{10}[f_\nu(i)/f_\nu(J)]$ 
versus sSFR for DustPedia and NIHAO. Both samples are here
limited to  face-on, disk galaxies. For NIHAO, we show the results for 
$\tau_{\rm clear}=2.5\,Myrs$ and $f_{\rm dust}=0.1$.}
\end{figure}

To make this comparison concrete, we consider how each galaxy's colors vary with 
axis ratio relative to its expected face-on color. For the simulated galaxies, 
we can actually just orient our system to see what the face-on colors of any 
given galaxy are, but for galaxies in the real universe we cannot do this. 
For real galaxies, we use the sSFRs measured for the DustPedia face-on galaxies to establish a relationship between measured sSFR and face-on colors; we 
then use this relationship for all DustPedia galaxies to infer what the face-on
color would have been.
We use sSFR because it is an inferred intrinsic property of the galaxy---intrinsic
in  the sense that it should not depend on axis ratio---that relates relative closely
to the flux ratios. This procedure assumes that
inferences from the DustPedia CIGALE fits do not depend on axis ratio, 

Figure \ref{fig:FitPlots} shows the colors and sSFRs of 
DustPedia face-on disk galaxies and 
NIHAO face-on disk galaxies (the latter under the free parameter choices of 
$\tau_{\rm clear}=2.5\,$Myrs and $f_{\rm dust}=0.1$).
We perform a linear regression of $\log_{10}[f_\nu(\rm FUV)/f_\nu(J)]$ on $\log_{10}({\rm sSFR})$ in each sample. This fit allows us to estimate the 
face-on color of any disk galaxies in either sample based on its sSFR.
We then define each 
galaxy's color residuals relative to these inferred face-on colors.

\begin{figure}
\begin{center}
\epsscale{1.17}
\plottwo{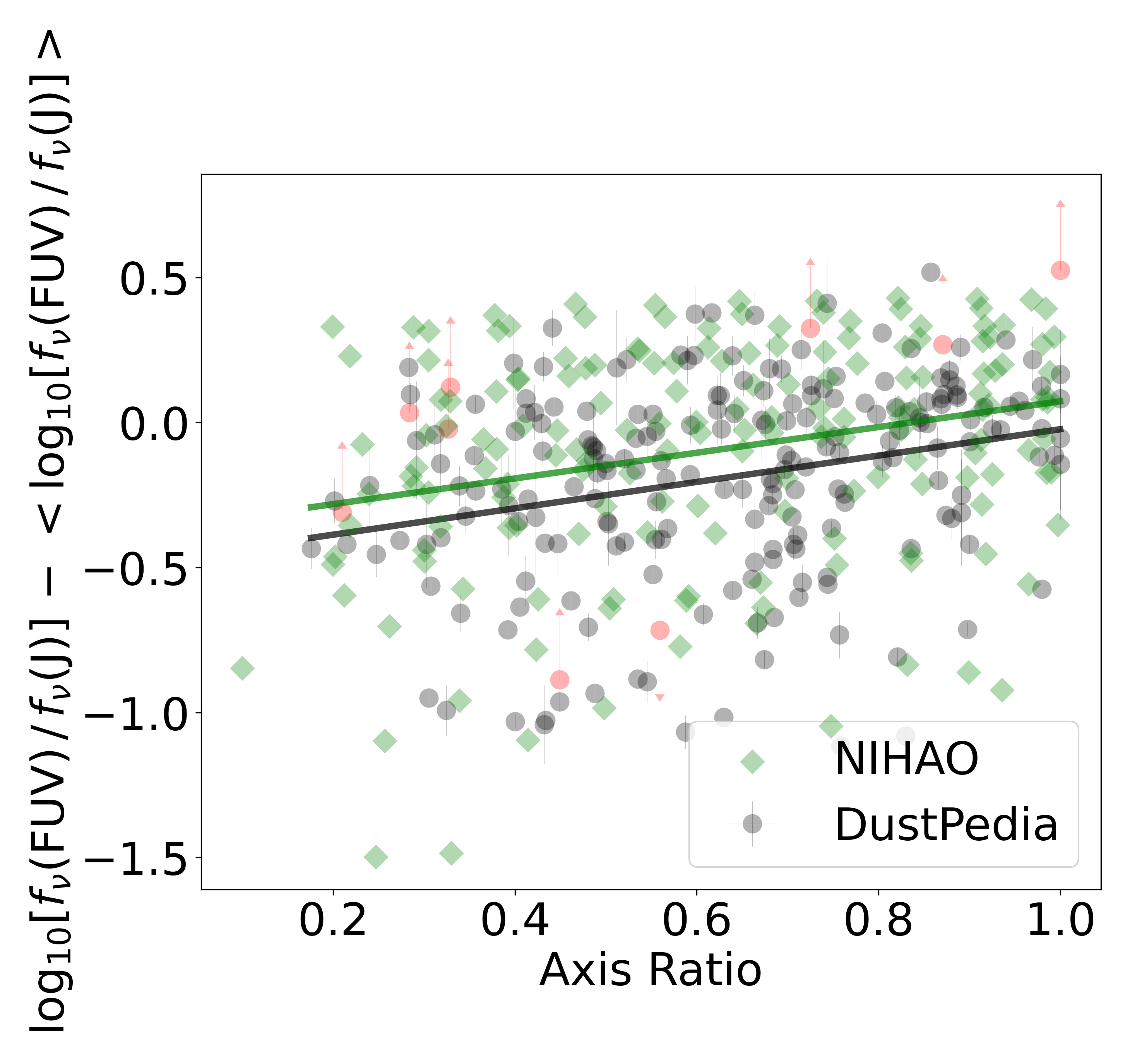}{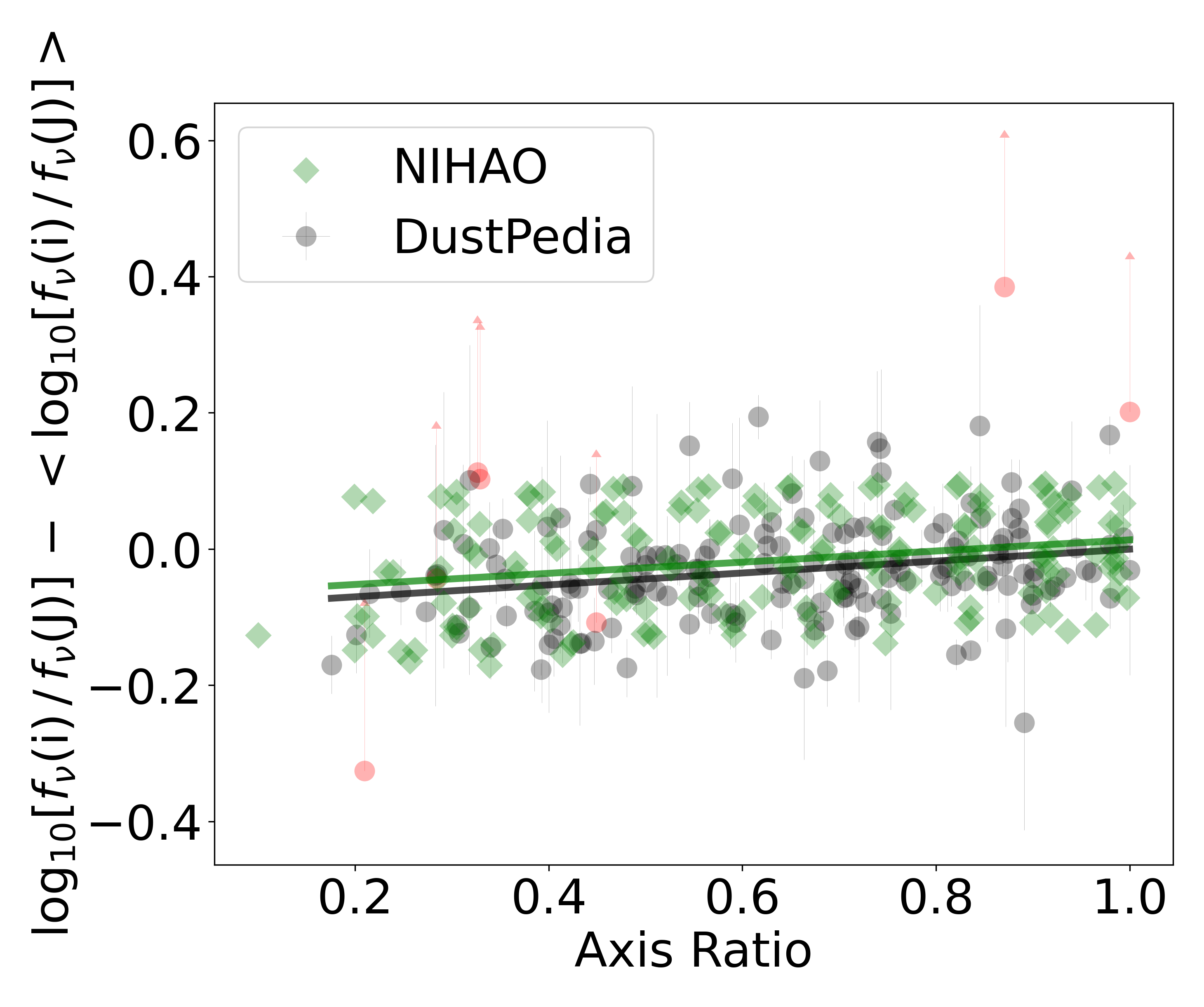}
\end{center}
\caption{\label{fig:ARCPs} \small 
Residuals relative to the expected face-on flux ratio given each galaxies
sSFR, as described fully in the text. The left panel shows the $\log_{10}[f_\nu(\rm FUV)/f_\nu(J)]$ color residual, and the right panel shows the $\log_{10}[f_\nu(i)/f_\nu(J)]$ color residual.
NIHAO galaxies use $\tau_{\rm clear}=2.5\,Myrs$ and $f_{\rm dust}=0.1$.}
\end{figure}

Figure \ref{fig:ARCPs} shows the distribution of these color residuals vs. axis ratio for NIHAO and DustPedia disk galaxies. These plots show the reddening effects measured by previous authors. We then perform linear regression of the residual $\log_{10}[f_\nu(\rm FUV)/f_\nu(J)]$ on axis ratio for these two populations. In Section \ref{Parameter Space}, we use the differences in slopes between DustPedia and NIHAO as a metric for constraining our free parameters.

\begin{figure}
\begin{center}
\epsscale{1.17}
\plottwo{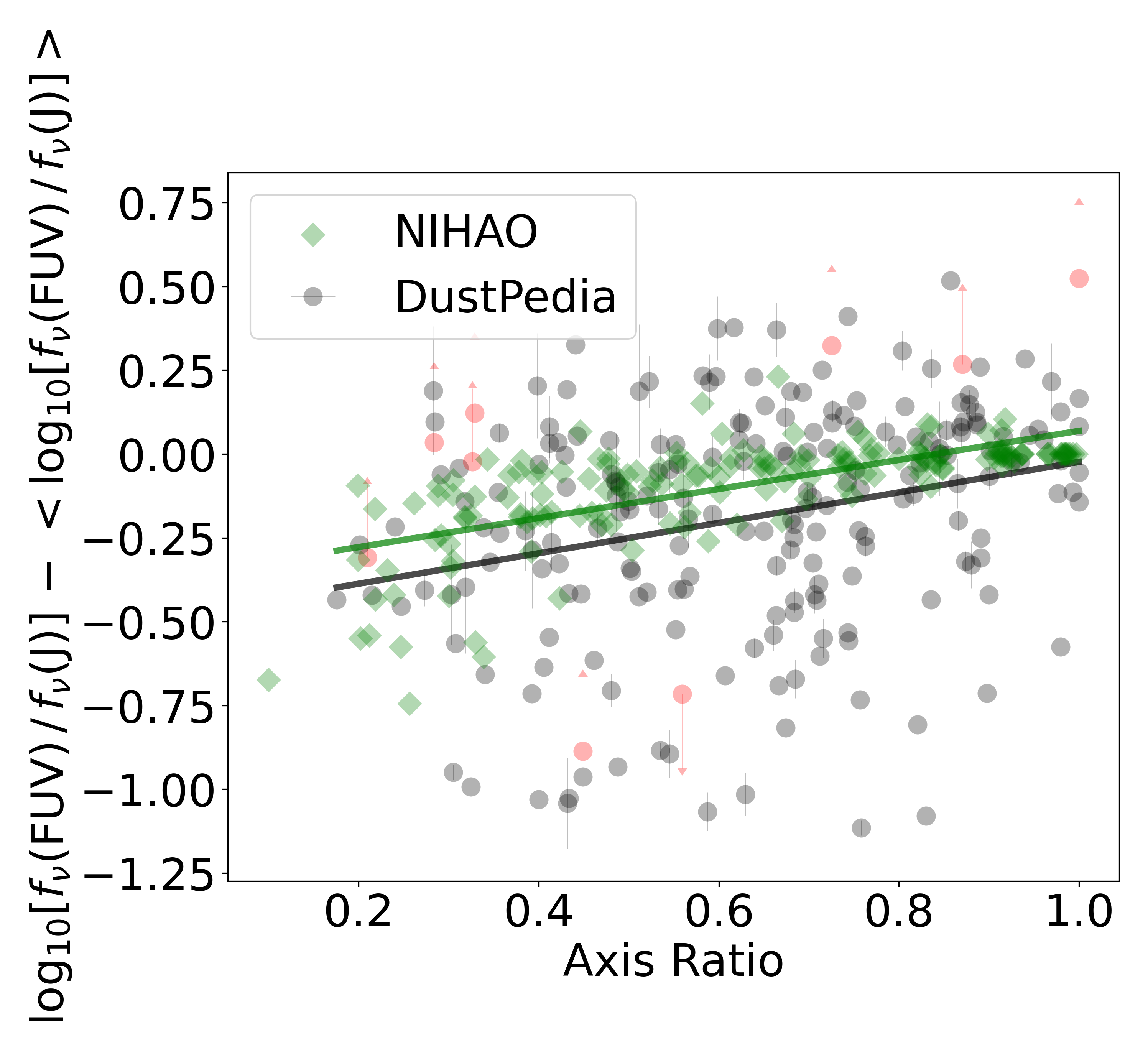}{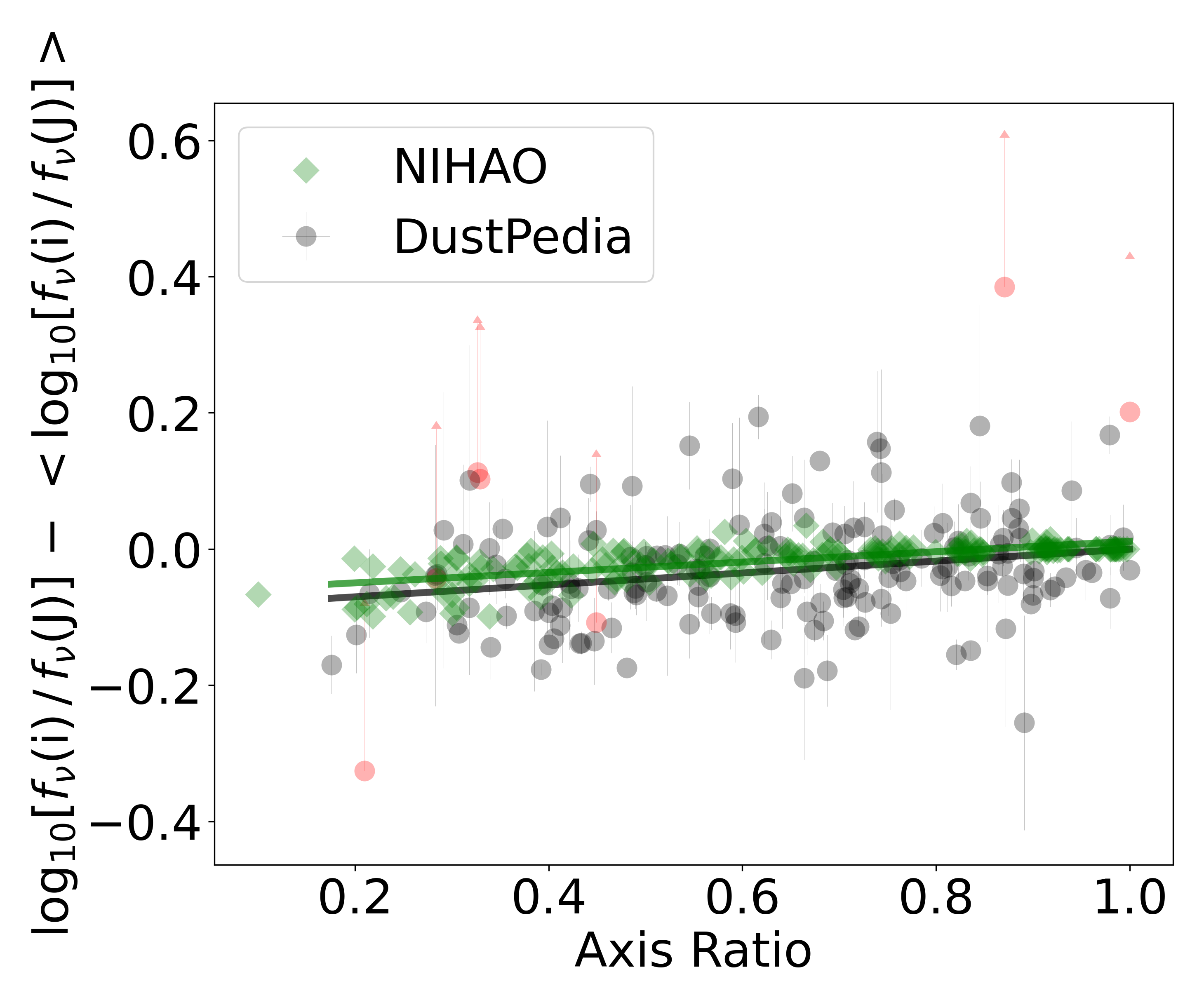}
\end{center}
\caption{\label{fig:separateARCPs} \small Similar to Figure \ref{fig:ARCPs},
but in this case the NIHAO galaxy residuals are relative to each galaxy's 
actual face-on colors. }
\end{figure}

The scatter around the lines in Figure \ref{fig:ARCPs} has some contribution
from an intrinsic scatter of the amount of reddening at a given axis 
ratio, and the scatter in the color at a given sSFR.
For the simulated galaxies, we can probe the intrinsic scatter around the 
axis ratio dependence more directly, since we know the actual face-on colors.
To probe this intrinsic scatter,
Figure \ref{fig:separateARCPs} subtracts the actual face-on 
colors of each NIHAO galaxy to determine fiducial colors. The most obvious 
difference between Figures \ref{fig:ARCPs} and \ref{fig:separateARCPs} is that, 
as we approach face-on orientations, we see the scatter in the NIHAO colors 
around the average go to zero. Notice however that the best-fitting slope is 
not likely to be affected by this choice, to the extent that
the sSFR of the NIHAO galaxies 
do not correlate strongly with their axis ratios in the simulated
observations.

\subsection{Parameter Space} \label{Parameter Space}

\begin{figure}
\begin{center}
\epsscale{1.17}
\includegraphics[width=0.7\textwidth]{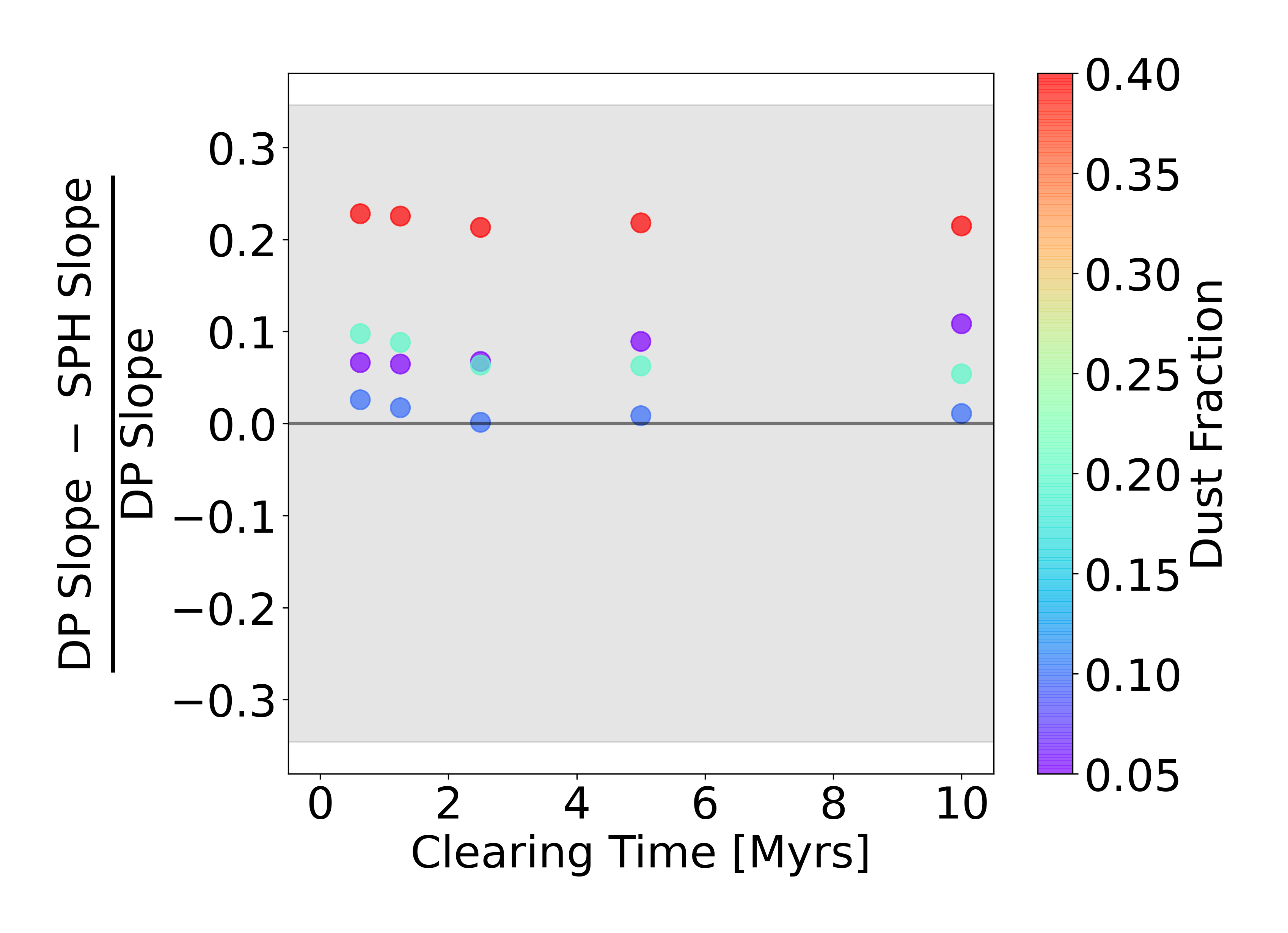}
\caption{\label{fig:BootstrapARCPSlopes}
\noindent Parameter search showing the distribution of the fractional difference in axis ratio color plot slope between DustPedia and the simulations. Parameter values are evenly separated in log space: $f_{\rm dust}=[0.05, 0.1, 0.2, 0.4]$ and $\tau_{\rm clear}=[0.625, 1.25, 2.5, 5, 10]\,$Myrs. The shaded region shows the bootstrap standard error averaged over all the data points, centered at $y=0$.}
\end{center}
\end{figure}

\begin{figure}
\makebox[\textwidth][c]{
\includegraphics[width=1.1\textwidth]{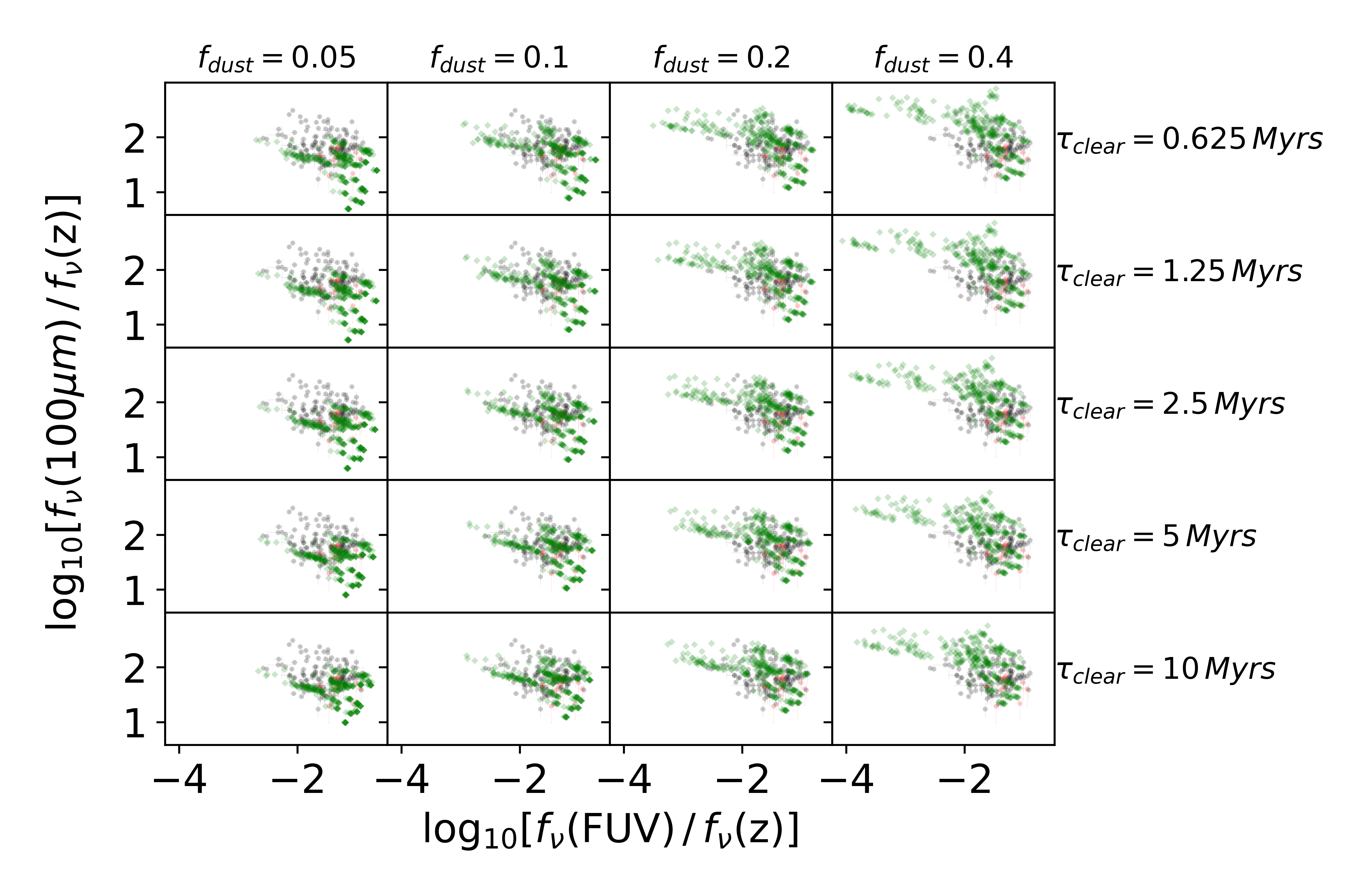}}%
\caption{\label{fig:pacs100MassCut}
Parameter space search showing the distribution of $\log_{10}[f_\nu(100 \mu m)/f_\nu(z)]$ vs. $\log_{10}[f_\nu(\rm FUV)/f_\nu(z)]$ for DustPedia and NIHAO. As in the previous plots, green diamonds represent NIHAO colors and black circles represent DustPedia colors. Our final parameter choices are $f_{\rm dust}=0.1$ and $\tau_{\rm clear}=2.5\,$Myrs.}
\end{figure}

\noindent To determine our free parameters, the clearing time and the dust fraction, 
we test a grid of clearing times 
(horizontal axis) and dust fractions (symbol colors). Parameter 
values are evenly separated in log space: 
$f_{\rm dust}=[0.05, 0.1, 0.2, 0.4]$ and 
$\tau_{\rm clear}=[0.625, 1.25, 2.5, 5, 10]\,$Myrs. 
Figure \ref{fig:BootstrapARCPSlopes} shows the fractional difference of 
the best-fitting $\log_{10}[f_\nu(\rm FUV)/f_\nu(J)]$ color residual vs. axis ratio slopes between
DustPedia and NIHAO disk galaxies corresponding to each grid point. The shaded region in this plot represents the standard error in the slope difference determined by bootstrap, averaged over all the data points, centered at $y=0$. For each $f_{\rm dust}$ and $\tau_{\rm clear}$, we calculate the bootstrap standard error by randomly sampling data points from NIHAO and DustPedia with replacement and repeatedly running the fitting procedures shown in Figures \ref{fig:FitPlots} and \ref{fig:ARCPs}. The standard deviation of the fractional slope differences among 100 iterations of sampling with replacement and fitting gives the bootstrap standard error for each combination of $f_{\rm dust}$ and $\tau_{\rm clear}$.
Figure \ref{fig:pacs100MassCut} shows the distribution of $\log_{10}[f_\nu(100\mu{\rm m})/f_\nu(z)]$ vs. $\log_{10}[f_\nu({\rm FUV})/f_\nu(z)]$ flux ratios for DustPedia and NIHAO corresponding to each grid point.

In Figure \ref{fig:BootstrapARCPSlopes}, even though the bootstrap standard error is large enough to make all the data points consistent with $y=0$, we can see that the $f_{\rm dust}=0.1$ fractional slope differences consistently show the best agreement with DustPedia. In Figure \ref{fig:pacs100MassCut}, we can also see that the $f_{\rm dust}=0.1$ column consistently shows the best agreement with DustPedia while different values of $\tau_{\rm clear}$ only have a small effect on the distribution of colors. Given this difficulty in constraining $\tau_{\rm clear}$, we settle on an intermediate value of $2.5\,$Myrs, leading to our final parameter choices of $f_{\rm dust}=0.1$ and $\tau_{\rm clear}=2.5\,$Myrs. 

Using the same values of $f_{\rm dust}$, we perform the same comparisons to the case of having no star-forming particles (all star particles are modeled with FSPS). In this case, the parameter space is only 1-dimensional and we find that $f_{\rm dust}=0.1$ has the best agreement with DustPedia galaxies. 

\subsection{Images} \label{Images}

\begin{figure}
\makebox[\textwidth][c]{
\includegraphics[width=1.75\textwidth]{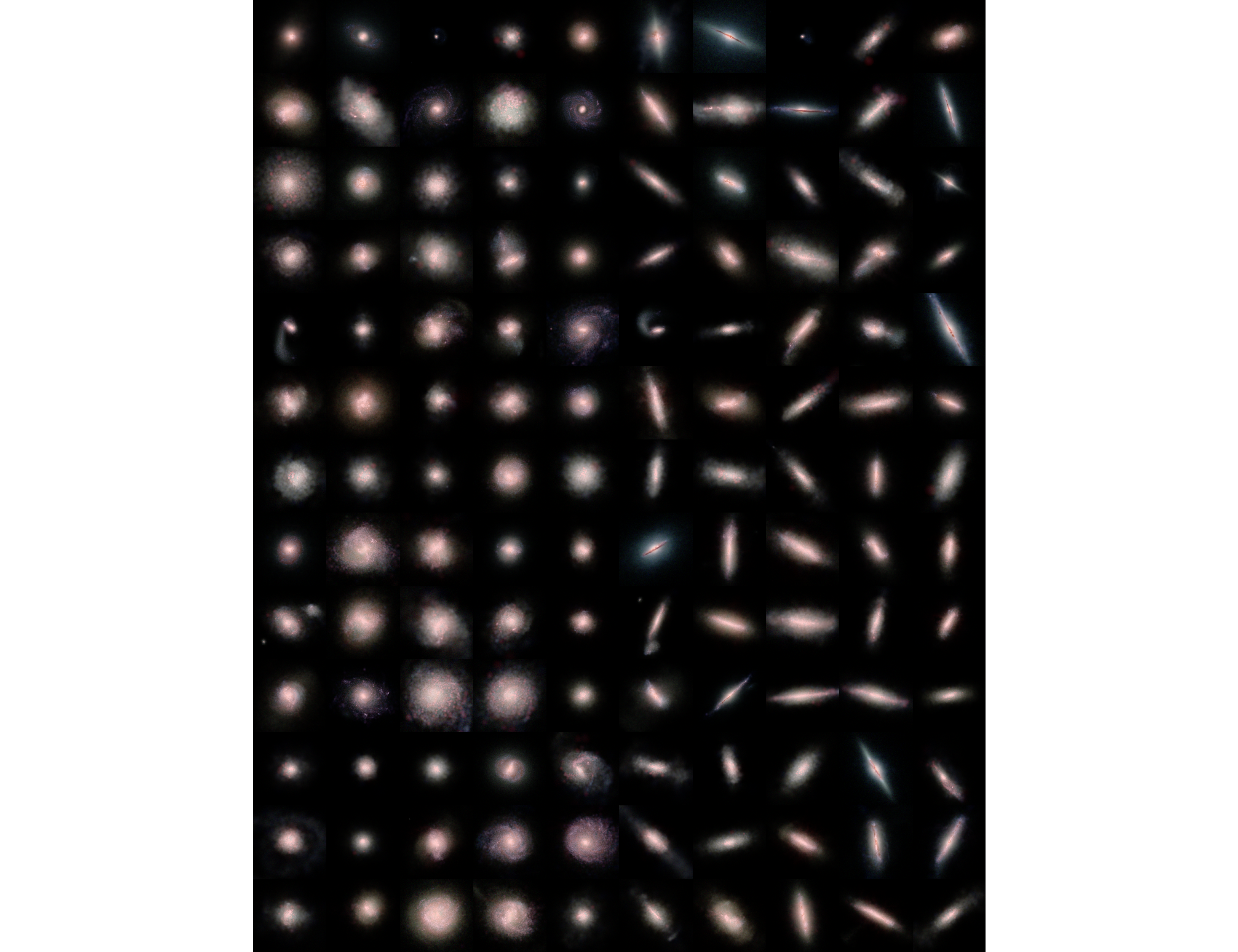}}%
\caption{\label{fig:faceEdgeImages}
Face-on (left 5 columns) and edge-on (right 5 columns) composite images of full galaxy sample.}
\end{figure}

\noindent Figure \ref{fig:faceEdgeImages} shows composite images of each NIHAO galaxy at their most face-on (left 5 columns) and most edge-on (right 5 columns) orientations. In order to keep the brightest pixels from dominating the images, each galaxy's fluxes are individually truncated, normalized, and scaled by an arcsinh stretch. We create the image by setting the red channel equal to $80\%$ z-band and $20\%$ W4-band, the green channel equal to $80\%$ r-band, and the blue channel equal to $72\%$ g-band and $8\%$ FUV-band. Note that these choices are purely cosmetic and have no impact on any of the quantitative results in this work.

\subsection{Color-color Plots} \label{Color-color Plots}

\begin{figure}
\begin{center}
\epsscale{1.17}
\plottwo{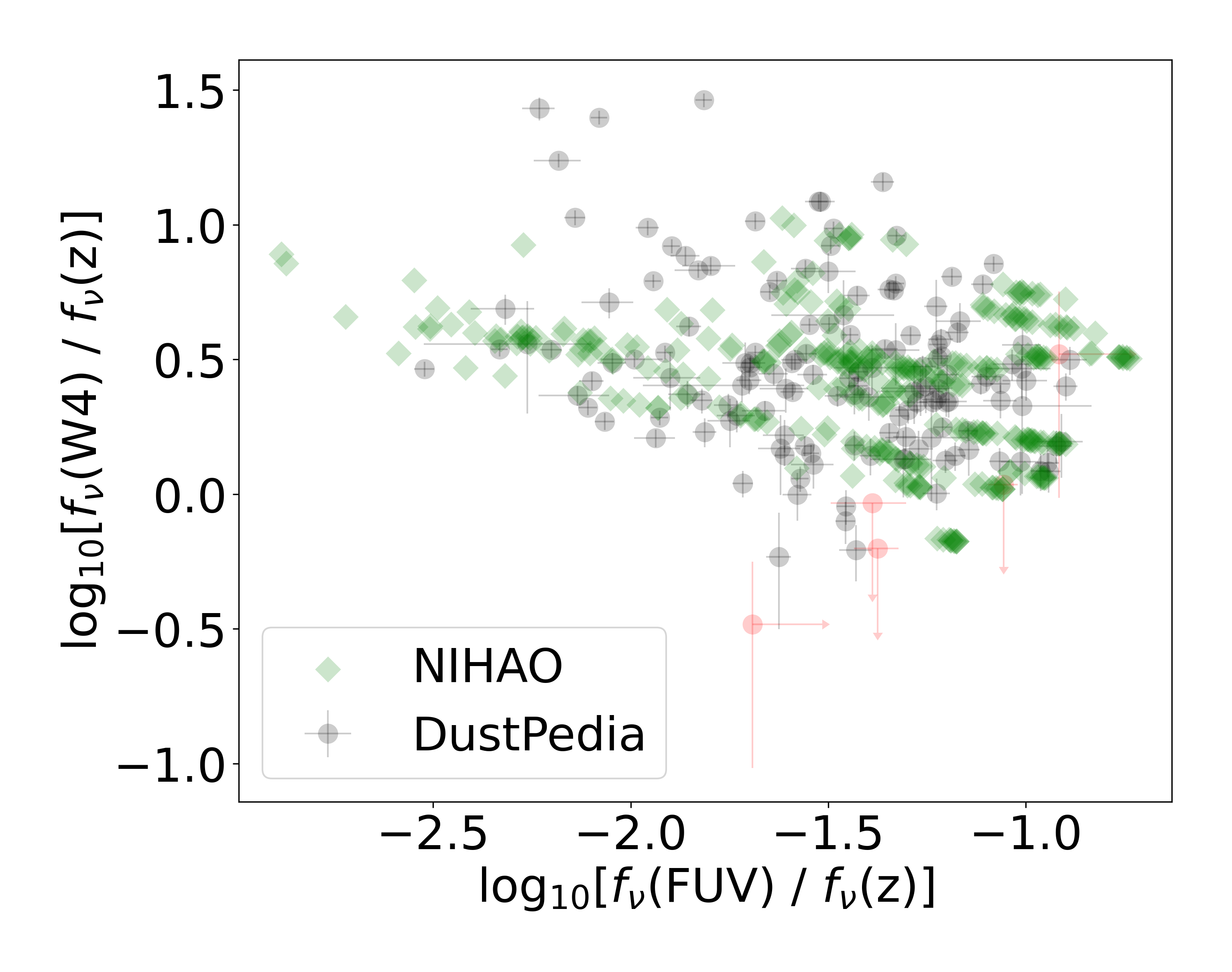}{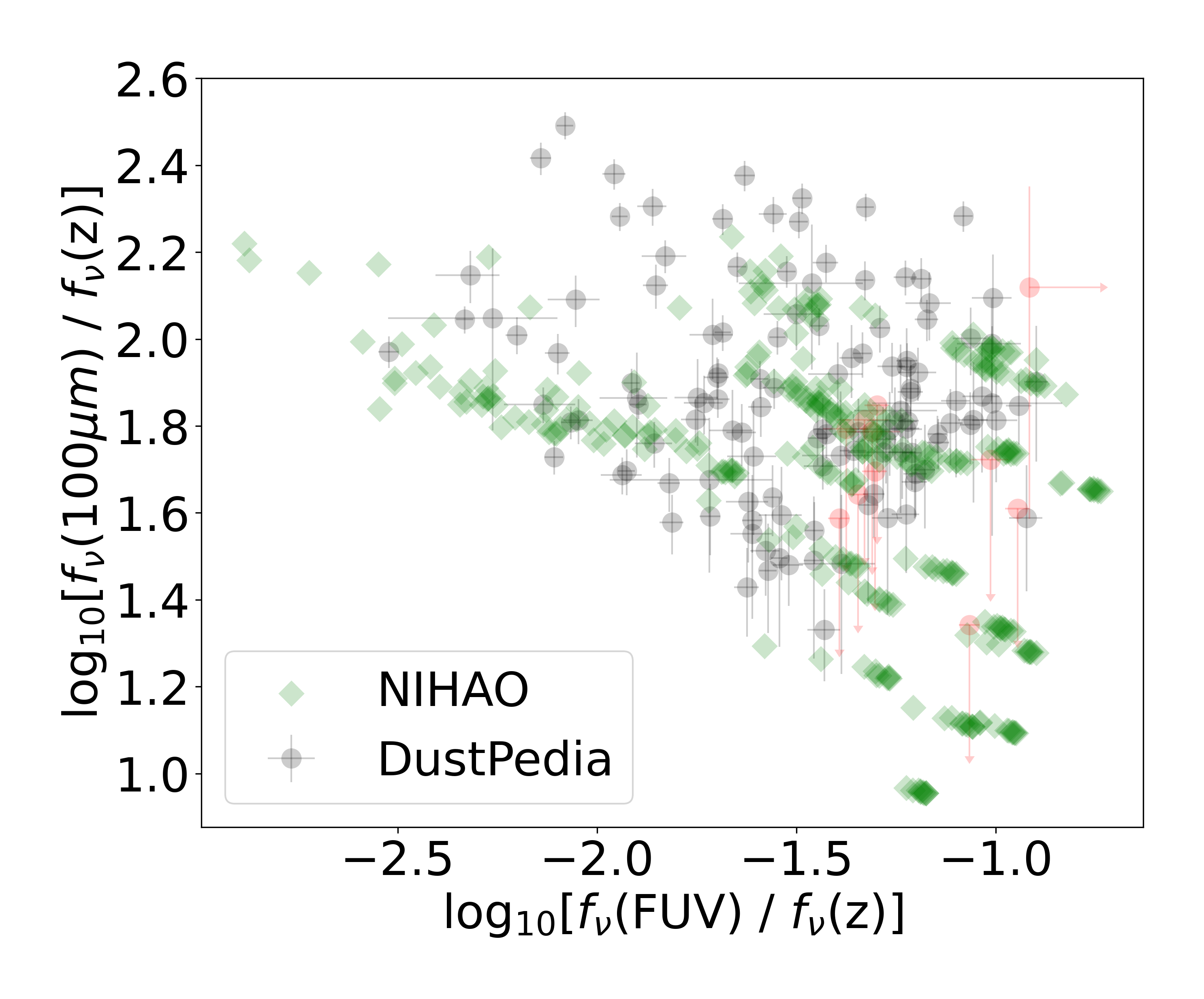}
\end{center}
\caption{\label{fig:Color-color Plots} \small 
Color-color plots for all orientations of NIHAO galaxies with $\tau_{\rm clear}=2.5\,$Myr and $f_{\rm dust}=0.1$, excluding galaxies with total stellar mass below $10^{9.5}\,M_{\odot}$.}
\end{figure}

\begin{figure}
\begin{center}
\epsscale{1.17}
\plottwo{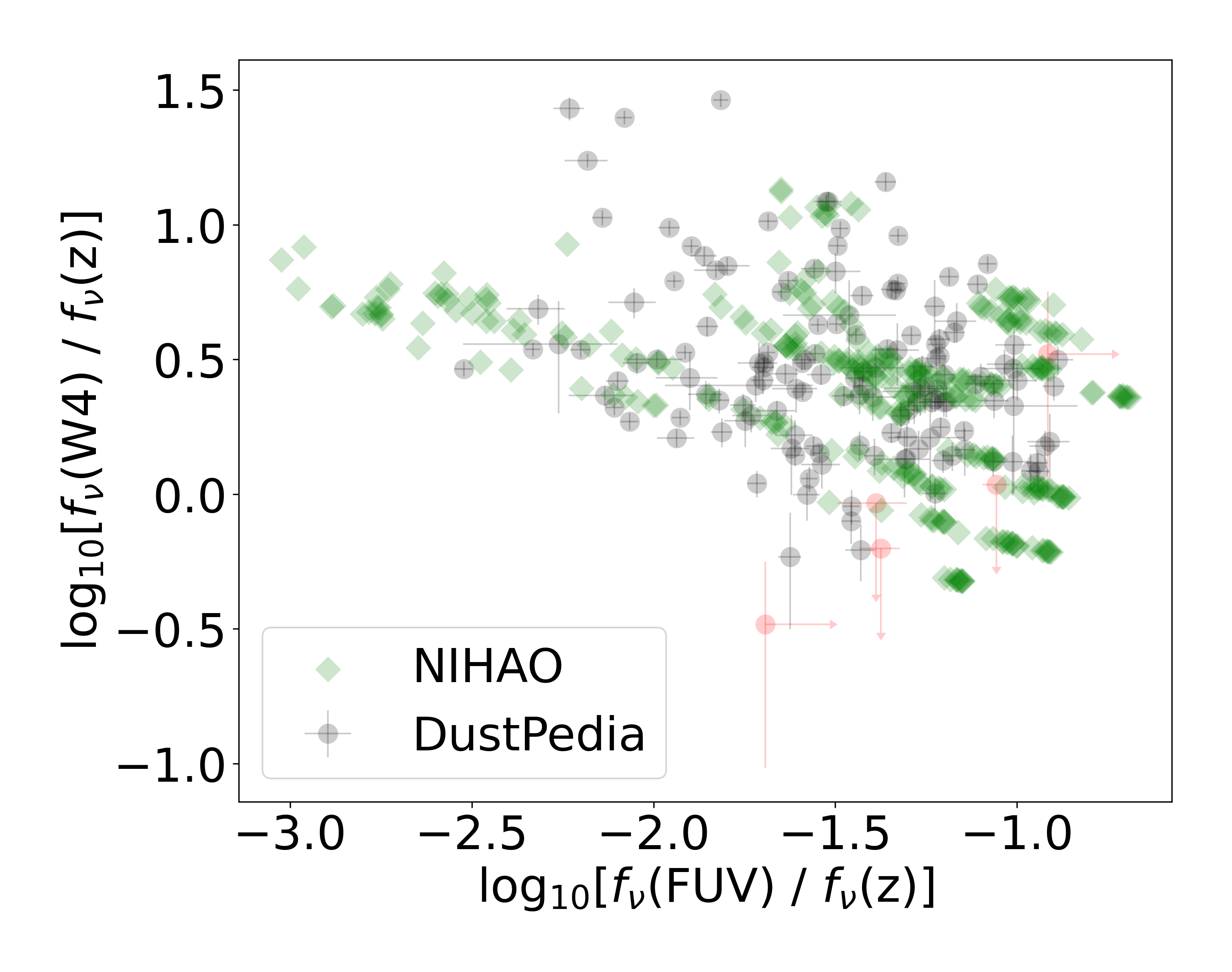}{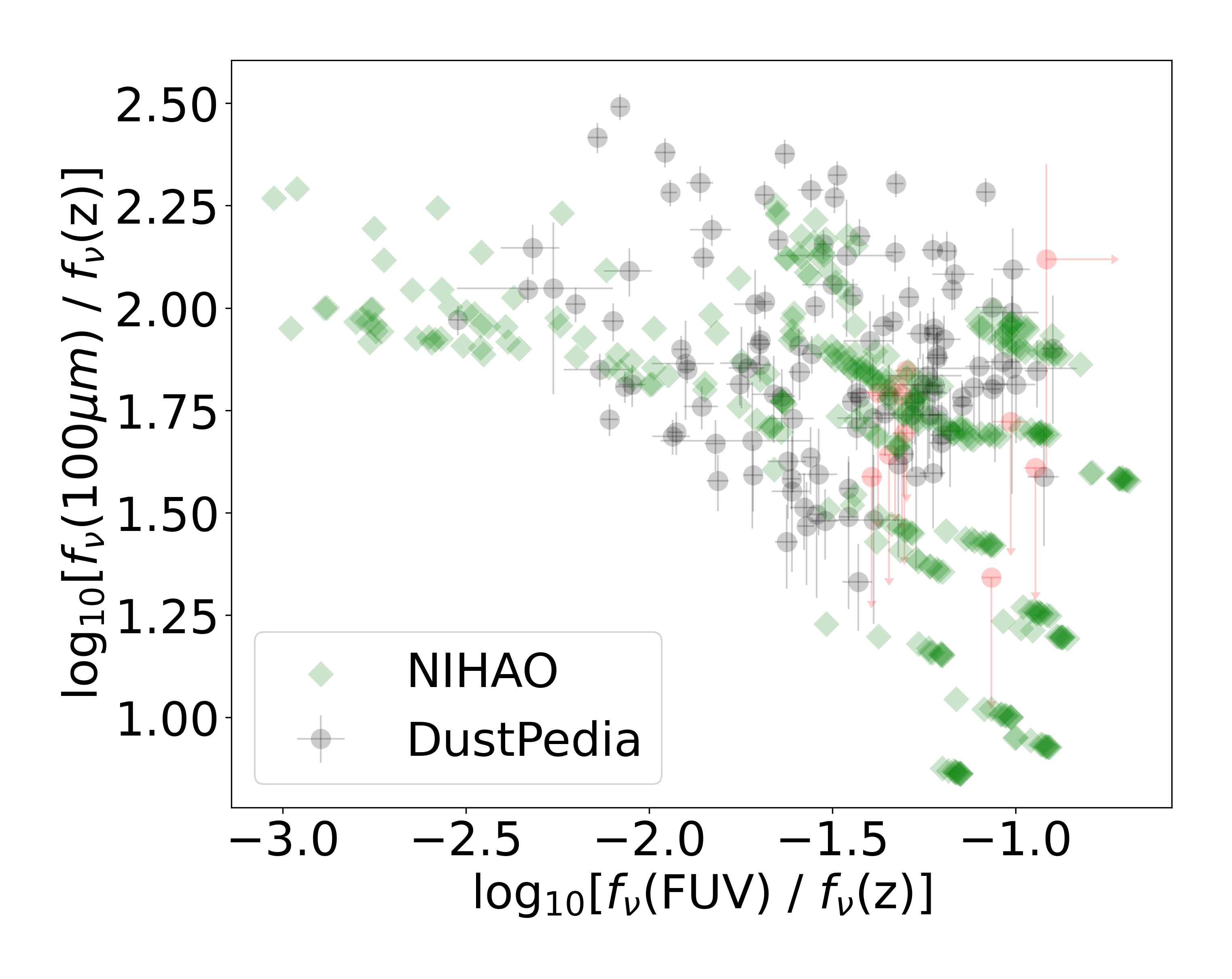}
\end{center}
\caption{\label{fig:noSF_Color-color Plots} \small 
Color-color plots for all orientations of NIHAO galaxies with no star-forming particles and $f_{\rm dust}=0.1$, excluding galaxies with total stellar mass below $10^{9.5}\,M_{\odot}$.}
\end{figure}

\noindent Looking at Figure \ref{fig:Color-color Plots}, we can see that our best-fitting simulated parameters qualitatively match the DustPedia galaxies' colors, with some outliers. While there are some galaxies that are deficient in both IR and FUV relative to optical light, we see that these discrepancies are even stronger when we don't include star-forming particles in our modeling, which is shown in Figure \ref{fig:noSF_Color-color Plots}. Whether the discrepancies in Figure \ref{fig:Color-color Plots} are due to limitations in the modeling of \HII{} regions, the diffuse dust model, or are inherent to the simulated galaxies is unclear.

\subsection{Attenuation curves} \label{Attenuation Curves}

\begin{figure}
\begin{center}
\epsscale{1.17}
\plottwo{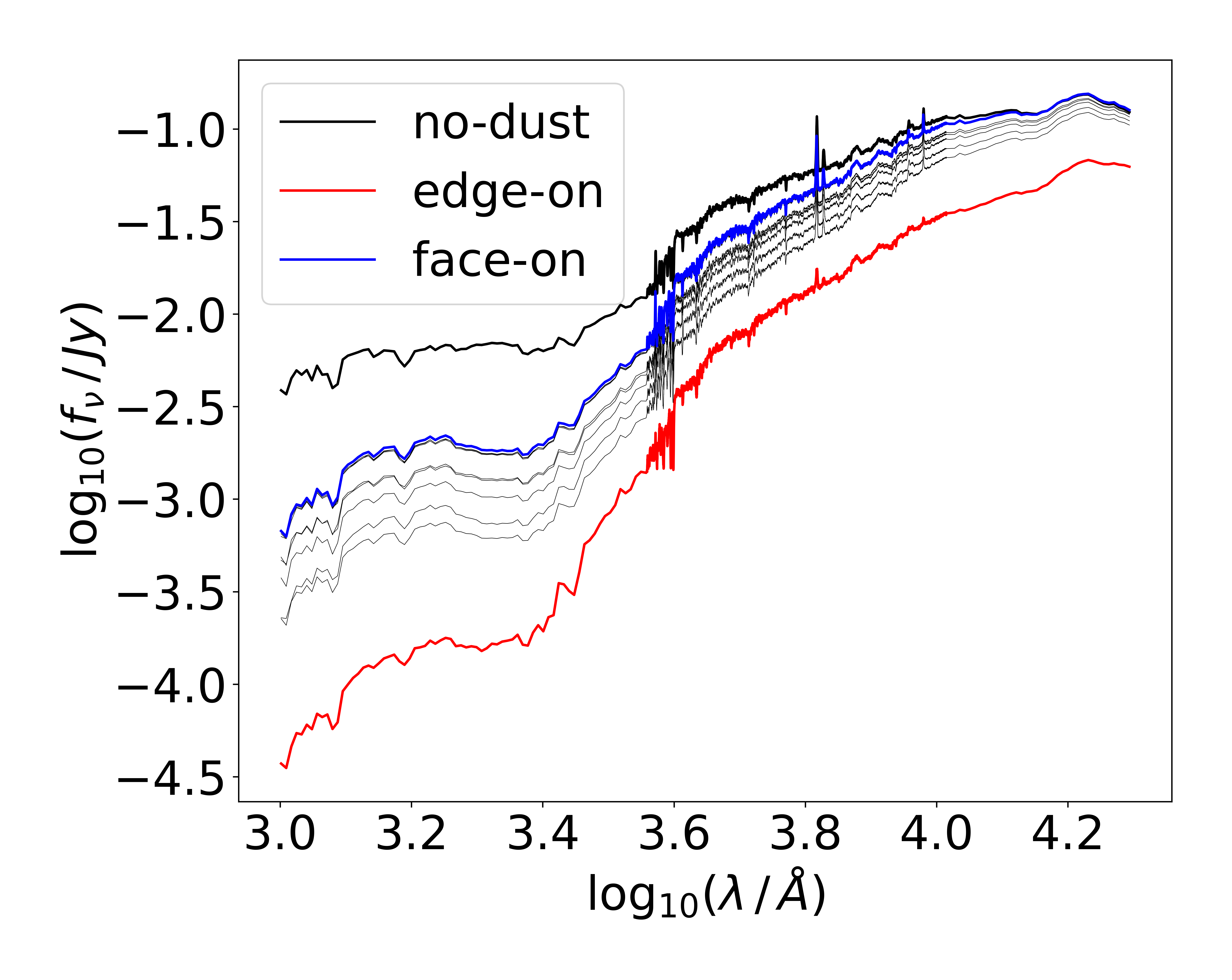}{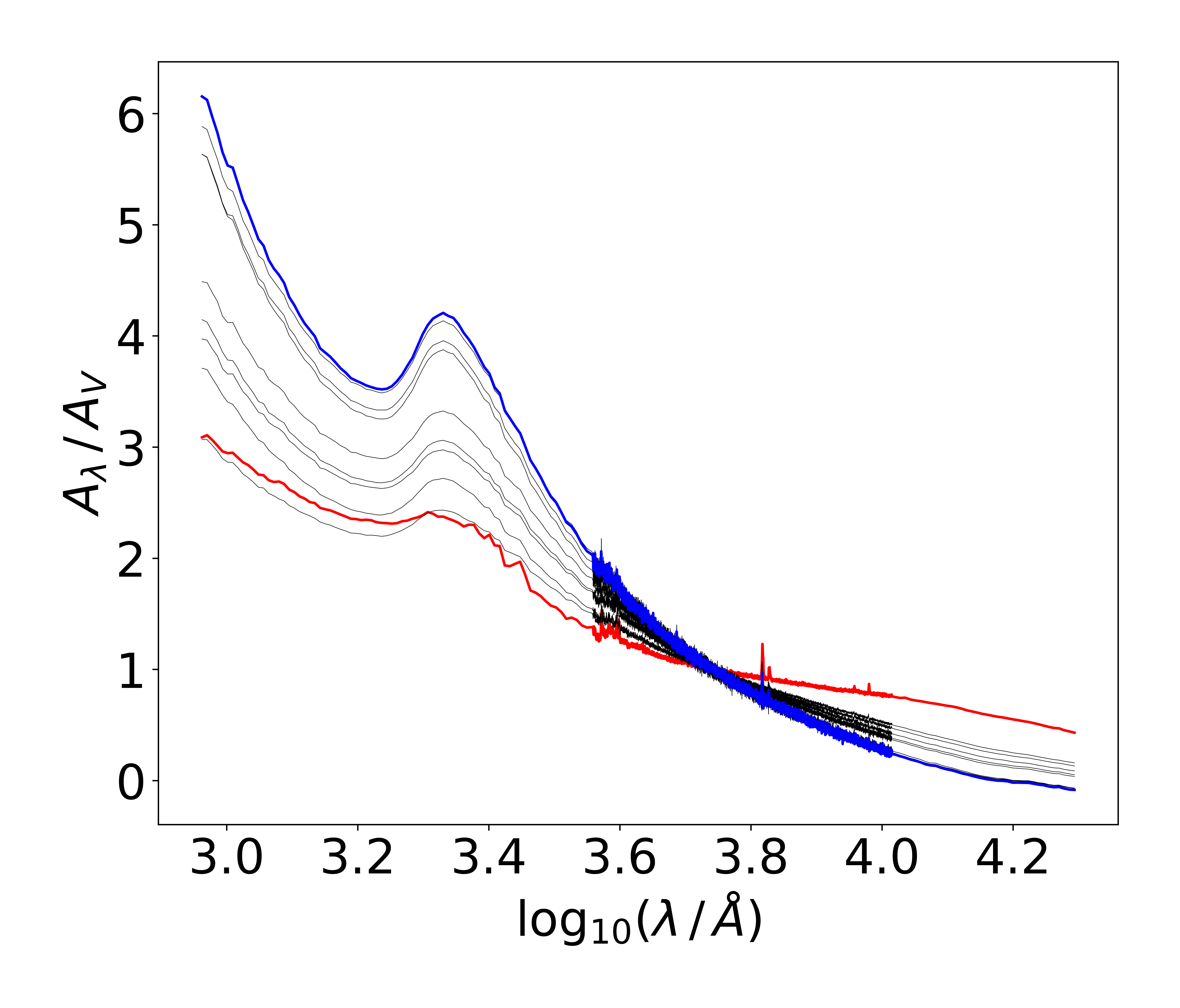}
\end{center}
\caption{\label{fig:SingleGalaxyAttenuation} \small 
SEDs (left) and normalized attenuation curves (right) for a single disk galaxy viewed from 10 different orientations. The black
curves show the results without dust, the blue curves show
the results for the face-on case, the red curves show the
results for the most edge-on case, and the grey curves show
intermediate axis ratios.}
\end{figure}

\begin{figure}
\begin{center}
\epsscale{1.17}
\plottwo{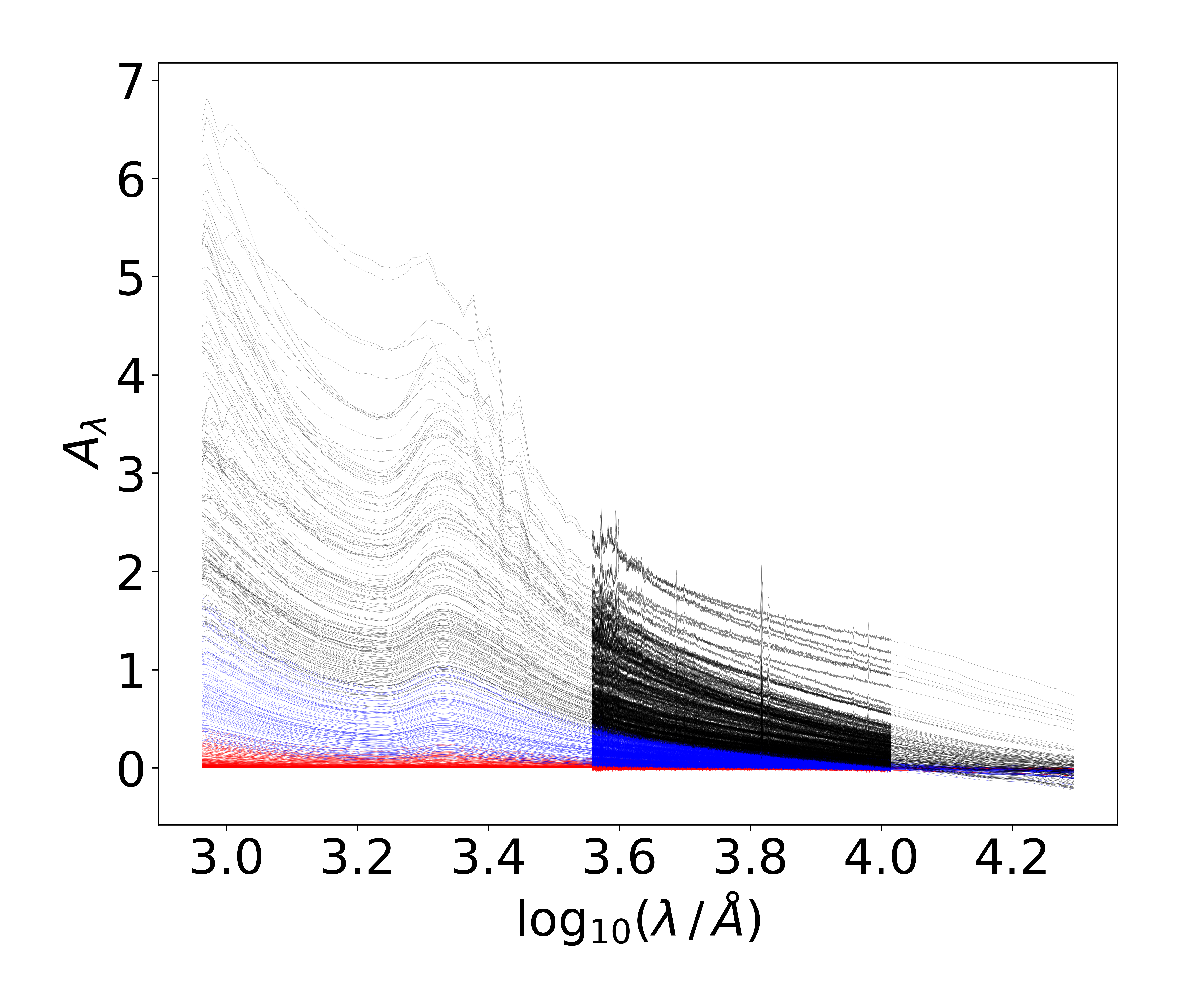}{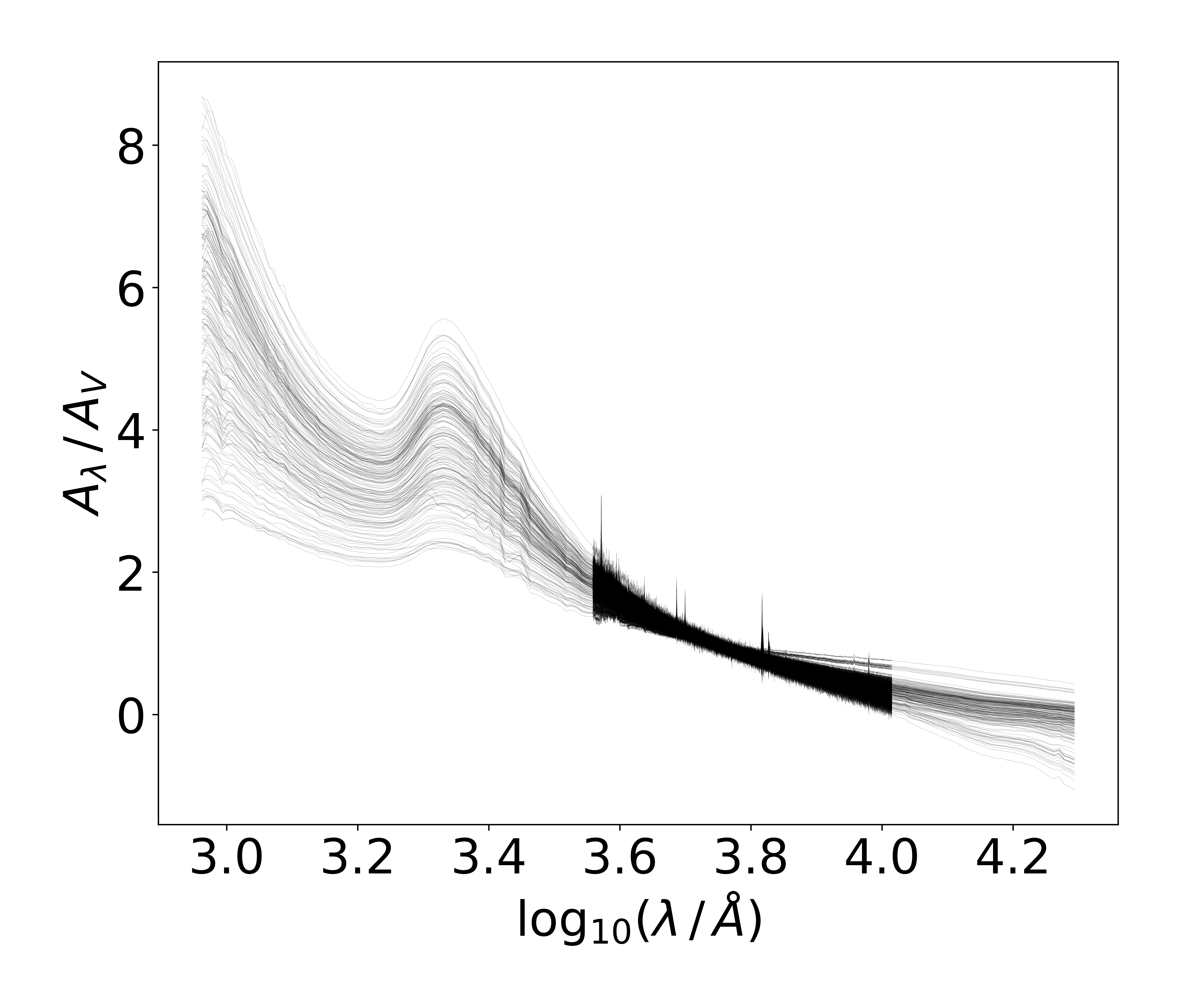}
\end{center}
\caption{\label{fig:Attenuation} \small 
Unnormalized attenuation curves (left) and attenuation curves normalized by $A_V$ (right) for the NIHAO galaxies with $\tau_{\rm clear}=2.5\,$Myr and $f_{\rm dust}=0.1$, excluding those with total stellar mass below $10^{9.5}\,M_{\odot}$. Red lines represent galaxies which have total stellar mass below $10^{9.5}\,M_{\odot}$, blue lines represent galaxies above this mass with $A_V<0.2$, and black lines represent galaxies above this mass and with $A_V>0.2$. 
The right hand panel only includes the
grey curves, i.e. those galaxies with $A_V>0.2$ mag.}
\end{figure}

\noindent To calculate the energy attenuated by dust, we run each simulated galaxy with no diffuse dust (effectively setting $f_{\rm dust}=0$), and no sub-grid dust from the photo-dissociation regions (PDRs) around young stars ($f_{\rm PDR}=0$). 
We then calculate the attenuation as:
\begin{equation}
   A(\lambda) = 2.5 \log_{10} \left(\frac{f_\nu(\lambda; {\rm ~no~dust})}
   {f_\nu(\lambda; {\rm ~dusty})}\right)
\end{equation}
For these simulations, this definition is suitable for $912\,$\AA\, to $2 \times 
10^{4}\,$\AA. However, even with $f_{\rm PDR}=0$, there is still a small 
amount of intrinsic attenuation and infrared emission coming from the 
star-forming stellar particles, and the emission contaminates 
$f_\nu(\lambda; {\rm ~no~dust})$ at longer wavelengths. 
We discussed in Section \ref{Source Model} that the attenuation 
at shorter wavelengths has a negligible effect 
on the  calculated attenuation curves. To compare the shapes of
the attenuation curves (i.e. the reddening), we will normalize 
them to the attenuation $A_V$ at 5500 \AA.

Figure \ref{fig:SingleGalaxyAttenuation} shows the spectra and 
normalized attenuation curves for a single simulated disk galaxy
viewed from 10 different orientations, with the face-on orientation 
shown in blue, the edge-on orientation in red, the no-dust spectrum in black, and intermediate orientations in grey. Looking at the normalized attenuation curves in this figure, we can see that there is significant variation in the shape of the attenuation curves at different viewing orientations. 

The left side of Figure \ref{fig:Attenuation} shows the 
attenuation curves of full simulated suite of NIHAO galaxies at 
all 10 orientations. Red lines correspond to galaxies with 
total stellar mass below $10^{9.5}\,M_{\odot}$, blue lines 
correspond to galaxies above this mass but with $A_V$ values 
less than 0.2, and grey lines correspond to galaxies above 
this mass and with $A_V$ values above 0.2. The right side 
of Figure \ref{fig:Attenuation} shows the normalized versions of the 
grey attenuation curves from the left side. Relative to Figure 
\ref{fig:SingleGalaxyAttenuation}, the normalized attenuation curves of Figure \ref{fig:Attenuation} show even more variation in shape.

The attenuation curves show a narrow  feature around 
$\lambda=6563$ \AA, the  wavelength of $H\alpha$ emission, 
and around other emission lines. 
These features result from differential attenuation of 
star forming regions  relative
to other regions.
In both the attenuated and unattenuated spectra, the 
\HII{} regions emit the same amount of $H\alpha$ emission. 
However, in the attenuated spectra, the $H\alpha$ emission is 
attenuated by the surrounding PDR while in the unattenuated 
spectra, the $H\alpha$ emission escapes the SF region freely. 
Because $H\alpha$ emission is preferentially located close to dust, in particular the sub-grid dust of the PDRs, we see a bump in the attenuation curve at $\lambda=6563$ \AA. Following this same line of reasoning, we can attribute the other consistent bumps and dips in the attenuation curves to other spectral features associated with either more or less dusty regions.

\subsection{Energy Balance} \label{Energy Balance}

\begin{figure}
\begin{center}
\epsscale{1.17}
\includegraphics[width=0.6\textwidth]{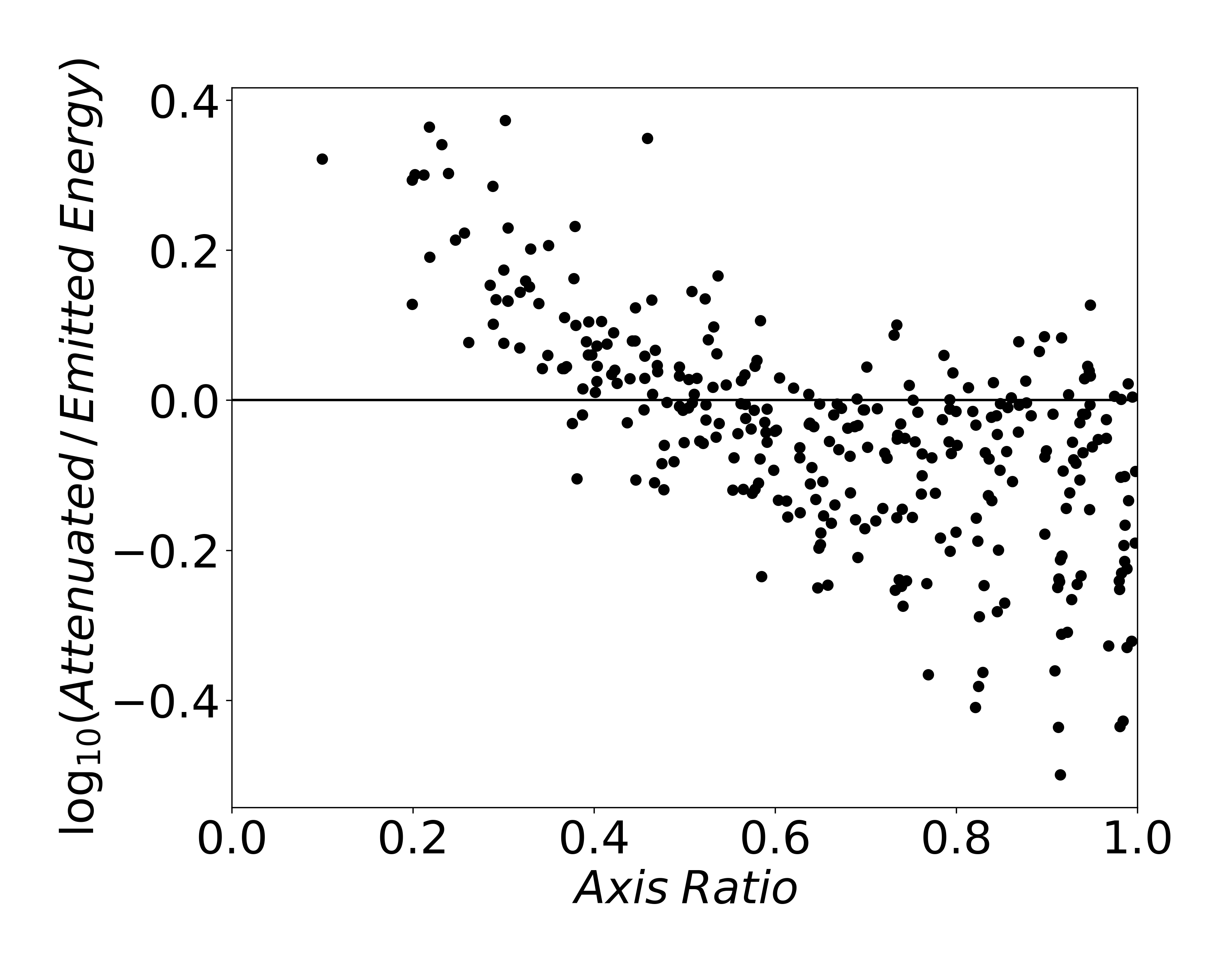}
\caption{\label{fig:EnergyBalance}
Energy balance plot for the NIHAO galaxies with $\tau_{\rm clear}=2.5\,$Myrs and $f_{\rm dust}=0.1$, excluding those with total stellar mass below $10^{9.5}\,M_{\odot}$.}
\end{center}
\end{figure}

\noindent Figure \ref{fig:EnergyBalance} shows the ratio of the energy attenuated and emitted by dust for simulated galaxies with total stellar mass above $10^{9.5}\,M_{\odot}$. Attenuated energy is calculated by integrating the difference between the no-dust and dusty spectra over wavelengths below $2\mu m$, and emitted energy is calculated by integrating the difference between the dusty and no-dust spectra over wavelengths above $2\mu m$.

Although the total energy (integrated over solid angle) attenuated by dust must equal the total energy reemitted by dust, this requirement does not hold when considering the light received by an observer in a particular direction. At low axis ratios, we can see that there is an excess of attenuated energy relative to emitted energy, and at large axis ratios we see the opposite effect. Qualitatively this is what we should expect, since, relative to a face-on galaxy, the stellar light from an edge-on galaxy will need to travel through more of the dust-filled ISM on average before reaching the observer. In general, we can see that there is strong deviation from perfect energy balance, which is denoted by the horizontal line. This result is in stark opposition to the assumptions made by commonly used SED modeling codes such as Prospector, CIGALE, and MAGPHYS, which impose strict energy balance between dust attenuation and emission. 

\section{Data Product} \label{Data Product}
\noindent To compliment this paper, we publicly release two data files containing our final mock photometry with model parameters $f_{\rm dust}=0.1$ and $f_{\rm PDR}=2.5$ Myrs, with the age smoothing procedure described in section \ref{Source Model}. All photometric quantities are calculated at a distance of 100 Mpcs. Both data files contain the following information for each simulated galaxy:

\begin{itemize}
  \item Total stellar mass
  \item Star-formation rate averaged over the most recent 100 Myrs
  \item Total dust mass
  \item Physical size
  \item Star-formation history (SFH) as a histogram of stellar mass as a function of age resolved by 200 age bins evenly separated in log space
  \item Chemical evolution history (CEH) as a 2D histogram of stellar mass as a function of age and metallicity resolved by 200x200 age and metallicity bins evenly separated in log space
\end{itemize}

\noindent We draw particular attention to the chemical evolution history (CEH), which can be used to recreate each simulated galaxy's photometry using your choice of stellar population synthesis (SPS) model. For each orientation of each galaxy, both data files contain the following information:

\begin{itemize}
  \item Axis ratio (b/a)
  \item $A_V$ 
\end{itemize}

\noindent The spatially integrated data file contains the following additional information for each orientation:

\begin{itemize}
  \item Total energy attenuated and emitted by dust (see Section \ref{Energy Balance} for more details)
  \item 20 spatially integrated broadband photometric bandpasses both with and without dust (FUV, NUV, u, g, r, i, z, J, H, K, W1, W2, W3, W4, $70 \mu$m, $100 \mu$m, $160 \mu$m, $250 \mu$m, $350 \mu$m, $500 \mu$m) 
  \item Spatially integrated high resolution spectra both with and without dust (see Section \ref{Wavelength Grids} for more details)
  \item Spatially integrated attenuation curves
\end{itemize}

\noindent The spatially resolved data file contains the following additional information for each orientation:

\begin{itemize}
  \item 20 spatially resolved (500$\times$500) broadband photometric bandpasses both with and without dust (FUV, NUV, u, g, r, i, z, J, H, K, W1, W2, W3, W4, $70 \mu$m, $100 \mu$m, $160 \mu$m, $250 \mu$m, $350 \mu$m, $500 \mu$m) 
\end{itemize}

\noindent The data product is publicly available at \url{https://github.com/ntf229/NIHAO-SKIRT-Catalog}.

\section{Summary} \label{Summary}

\noindent By utilizing the radiative transfer software SKIRT, we have created a mock catalog of realistic photometry from the NIHAO galaxy simulation suite. Statistical comparison to observed galactic flux ratios in the local universe has motivated the implementation of sub-grid recipes that mitigate limitations in the temporal and spatial resolution of the simulations (Section \ref{Source Model}). Using this mock catalog, we have shown that commonly made simplifying assumptions in SED modeling do not hold true when we spatially resolve the interactions between starlight and cosmic dust. Of particular importance, we find that the assumption of energy balance between dust attenuation and emission along an observer's line of sight can be violated by up to a factor of 3 (Section \ref{Energy Balance}). We publicly release our final mock photometry as a data product, which also contains physical properties about each simulated galaxy and viewing orientation (Section \ref{Data Product}). Future work will be dedicated to using this mock catalog as a ground truth to determine the accuracy of SED modeling inferences under a variety of modeling choices and assumptions. 

\clearpage

\bibliography{refs}{}
\bibliographystyle{aasjournal}

\clearpage

\end{document}